\documentclass{article}

\usepackage{arxiv}

\usepackage{amsmath}
\usepackage[utf8]{inputenc} 
\usepackage[T1]{fontenc}    
\usepackage{hyperref}       
\usepackage{url}            
\usepackage{booktabs}       
\usepackage{amsfonts}       
\usepackage{nicefrac}       
\usepackage{microtype}      
\usepackage{lipsum}		    
\usepackage{graphicx}
\usepackage{natbib}
\usepackage{doi}
\usepackage{caption}
\usepackage{subcaption}
\usepackage{makecell}
\usepackage{multirow}
\usepackage[linesnumbered,ruled]{algorithm2e}

\usepackage{tikz}
\usetikzlibrary{shapes.callouts}
\usetikzlibrary{decorations.text}
\usetikzlibrary{arrows,calc,chains,positioning,shapes.callouts}
\usepackage{tikzpeople}
\usetikzlibrary{bayesnet, positioning, calc}

\title{Joint Inference of Diffusion and Structure in Partially Observed Social Networks Using Coupled Matrix Factorization}

\author{%
    Maryam Ramezani \\
    Computer Engineering Department \\
    Sharif University of Technology \\
    \texttt{maryam.ramezani@sharif.edu} \\
    \And
    Aryan Ahadinia \\
    Computer Engineering Department \\
    Sharif University of Technology \\
    \texttt{aryan.ahadinia@sharif.edu} \\
    \And
    Amirmohammad Ziaei \\
    Computer Engineering Department \\
    Sharif University of Technology \\
    \texttt{ziaei@ce.sharif.edu} \\
    \And
    Hamid R. Rabiee \\
    Computer Engineering Department \\
    Sharif University of Technology \\
    \texttt{rabiee@sharif.edu} \\
}

\date{}

\renewcommand{\headeright}{}
\renewcommand{\undertitle}{}
\renewcommand{\shorttitle}{Joint Inference of Diffusion and Structure in Partially Observed Social Networks Using Coupled Matrix Factorization}

\hypersetup{
pdftitle={Joint Inference of Diffusion and Structure in Partially Observed Social Networks Using Coupled Matrix Factorization},
pdfsubject={cs.SL},
pdfauthor={Maryam Ramezani, Aryan Ahadinia, Amirmohammad Ziaei, and Hamid R. Rabiee},
pdfkeywords={Social Network, Information Diffusion, Partially Observed Data, Network Structure, Matrix Factorization, Link Prediction, Cascade Completion, Bayesian Computation, Coupled Matrix Factorization},
}

\begin{document}
\maketitle

\begin{abstract}

Access to complete data in large-scale networks is often infeasible. Therefore, the problem of missing data is a crucial and unavoidable issue in the analysis and modeling of real-world social networks. However, most of the research on different aspects of social networks does not consider this limitation. One effective way to solve this problem is to recover the missing data as a pre-processing step. In this paper, a model is learned from partially observed data to infer unobserved diffusion and structure networks. To jointly discover omitted diffusion activities and hidden network structures, we develop a probabilistic generative model called "DiffStru." The interrelations among links of nodes and cascade processes are utilized in the proposed method via learning coupled with low-dimensional latent factors. Besides inferring unseen data, latent factors such as community detection may also aid in network classification problems. We tested different missing data scenarios on simulated independent cascades over LFR networks and real datasets, including Twitter and Memtracker. Experiments on these synthetic and real-world datasets show that the proposed method successfully detects invisible social behaviors, predicts links, and identifies latent features.

\end{abstract}

\keywords{
Social Network \and
Information Diffusion \and
Partially Observed Data \and
Network Structure \and
Matrix Factorization \and  
Link Prediction \and
Cascade Completion \and
Bayesian Computation \and
Coupled Matrix Factorization
}

\section{Introduction}\label{sec:introduction}
Social networks are essential platforms for the interaction of people through an explicit link by following each other and an implicit connection by sharing information. The widespread use of social networks by increasing the number of users, the large number of interactions between them, and the amount of information propagation over these networks have led to a line of research focused on analyzing and modeling these networks. The target audience for the results of this research are either:
\begin{enumerate}
	\item The owners of these platforms (e.g., Twitter, Instagram) 
	\item The third-party enterprises with customers who use these platforms (e.g., advertising companies, product providers, news analysts). 
\end{enumerate}

This paper focuses on solving the missing data problem for the latter community. 
In the context of large-scale social media, data collection is a massive, expensive, and time-consuming task for third-party companies. Typically, social datasets are provided for a limited period and include a subset of users for specific applications. In practice, having access to the complete data of a network is impossible even for a short period, and we often observe a partial subset of social data because of the following reasons:

\begin{enumerate}
	\item API call restrictions: Most social network platforms provide public API for data access in well-defined formats and automatically trigger a rate-limiting mechanism. For example, Twitter, Instagram, and Facebook impose a rate limit on getting a user's list of following/followers or posts,
	\item Rate-limit for web crawler: Websites usually set a limited number of requests for fetching data per IP address. When a crawler scraps the site, a temporary ban occurs when its request rate exceeds the predefined limit,
	\item Sampling technique: Due to the large volume and variety of data in a network, sampling methods are used for prioritization. Depending on the sampling method, missing data arise through the collection process,
	\item Protecting privacy: Statistics have shown that social network users' private accounts and rates of private activities are steadily on the rise \cite{vesdapunt2015identifying}. Since web crawlers do not have access to the information of the private accounts, the collected dataset does not represent the complete data.
\end{enumerate}

Although much research has been done on social network aspects, most do not consider the problem of missing data. Hence, the completeness assumption in those social network methods affects their performance on actual data. To alleviate this problem, some methods utilize a pre-processing algorithm on the input datasets to obtain a set with the least missing data to compensate for their completeness assumption. The main question is: How can we apply the existing methods to the original crawled data of social networks that contain missing data? In this paper, we try to provide an answer to this question for a specific type of data (graph-structured data with diffusion information) by proposing a probabilistic generative method called DiffStru. \\	
A network can be modeled by a graph including users and their interactions. The activity of users {in} time is another data representing a phenomenon over a network named diffusion. Diffusion is usually known as a set of propagation processes called a cascade. A cascade transmits information from one user to another connected user in a network. A user is said to be infected in a cascade if it receives republished information from another user. In this context, infection time is the time of user activity in publishing or republishing information in a cascade. It is good to note that in most social platforms, the name of nodes and their infection time is the only accessible data from cascades and other additional data, such as the infection path and the trace of who infected who is unknown. While the cascade graph is not accessible, the partial sets of nodes and infection times in each cascade are available. These concepts in a social network like Twitter are: users having an account on Twitter can be modeled as nodes of a graph with directed links expressing following and follower relations between them. When a user tweets a piece of information, her/his followers will receive it, and the sequence of the tweet's retweets will establish a cascade. A collection of cascades for different tweets forms diffusion. 

\begin{figure}[t]
    \centering
    \begin{subfigure}{0.4\textwidth}
        \centering\includegraphics[width=1\textwidth]{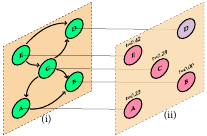}
        \caption{Groundtruth}
    \end{subfigure}
    \begin{subfigure}{0.4\textwidth}
        \centering\includegraphics[width=1\textwidth]{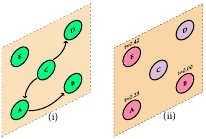}
        \caption{Observed}
    \end{subfigure}
    \caption
    {
        {Our problem at a glance. A structural network is described by frame (i): If $A$ follows $B$, there is a link from $A$ to $B$. Frame (ii) focuses on diffusion behaviors, where pink-colored nodes indicate infected nodes with their time of infection. Part (b) contains the same frames as part (a), with missing data in each frame indicating partial observations. This paper aims to extract the complete data (a) by combining structural information (frame (i)) with diffusion information (frame (ii)) in the observed mode (b).}
    }
    \label{fig:problem}
\end{figure}

Even if we consider a set of nodes in data collection and obtain the links between them for a time interval, we have some unobserved links due to the above-mentioned limitations. On the other hand, unobserved information in diffusion data appears as missing activity of nodes in each cascade. Fig. \eqref{fig:problem} illustrates the problem we are trying to solve. Given a limited period with a partially observed network graph structure and diffusion information containing the same set of users, our goal is to complete the missing data.
\\We make the following assumptions:
\begin{enumerate}
	\item  We assume that the data are collected over a limited period, such that the network structure remains static (no new users join or leave the network), and a timestamp for the creation of the link is not available during that period.
	In addition, this assumption implies that the diffusion information collected during crawling is based on a static structure. Thus, the network links are also static, and the data is fixed during interactions. On the other hand, a cascade has a limited lifespan. It spreads over hours to days, which means the sequence from beginning to end is usually fixed within a specific data collection interval.
	\item We are not interested in hidden nodes with no global or local signals in structure or diffusion data. However, we can somewhat handle users with private accounts. Although we cannot observe their profiles and activities, we can obtain their interactions from the list of following/followers of their visible neighbors. Therefore, the one-step neighbors of visible users are not hidden, while we are not engaged in any hidden neighbors of private users.
	\item We consider that the source and mediators of an information dissemination process are internal network factors and any external agents are ignored.
\end{enumerate}

As mentioned before, incomplete data in the collected set from a social network is inevitable, and missing data can significantly affect the difference between the output of methods and what happens in the real world \cite{Phantom2016}. Considering the missing data in two levels of structure and diffusion simultaneously makes the data inference challenging. Here, we try to make the inference tractable by selecting the appropriate distributions on the data. We present a novel generative model for jointly inferring the partial network structure and the information diffusion. \\
The contributions of this paper are as follows:
\begin{itemize}
	\item Tackling the problem of missing data \textbf{both in diffusion and structure} in real-world social datasets by proposing a new probabilistic generative model to jointly discover the hidden links of network structure and omitted diffusion activities. 
	\item Investigating structure and diffusion joint properties via probabilistic matrix factorization.
	\item Inferring the missing links of a network structure and the missing activities of nodes when a partial graph and a set of cascades with missing data are observable. 
	\item Inferring the diffusion behavior of users even when the user page is private and we have no information about his activities.
	\item Demonstrating that the low-dimensional representations for characteristics of users and cascades during the inference of DiffStru can be widely used in embedding and classification problems.
\end{itemize}

The rest of the paper is organized as follows: First, in Section \eqref{sec:background}, we present a review of social research related to partially observed data that motivates the paper. In Section \eqref{sec:related work}, we review the
related work, while Section \eqref{sec:problem statement} contains the paper notations, definitions, and problem statement. We present the proposed model in Section \eqref{sec:proposed solution}. In Section \eqref{sec:numerical and simulation results}, we examine the empirical results on synthetic and real-world datasets. Finally, we present the conclusions by discussing future work in Section \eqref{sec:conclusion}.

\section{Historical Background and Motivation}\label{sec:background}
Partially observed data has received enormous attention in previous studies of social networks. In the scope of this paper, social data is divided into two classes: \textbf{structure} networks (nodes and links between them) and \textbf{diffusion} (cascades propagation over a network). Missing data can occur in each of these two classes. 

Regarding history, missing data was first raised for structural network data, a long-standing research track known as link prediction. The main goal of this track is to find the missing links using observed interactions without considering diffusion information. Over the years, different assumptions and methods have been employed to solve the link prediction problem. 

Over time, due to the focus on the correlation between the diffusion process and network structure, both data classes have been considered together. In contrast, missing data is assumed at either the diffusion or structure class (but not both). 
For missing data at the diffusion class, one of the first attempts to analyze partially observed cascades was \cite{CorrectingLeskovec2011}. As long as the structural network is complete, the model attempts to estimate the properties of a whole cascade by using incomplete missed cascade samples.
In missing data at structure class, \cite{Phantom2016} examines the characteristics of the diffusion process over partial network structures. It attempts to infer the truth of cascade broadcasts occurring in the underlying network.

In the later years, different studies were published on the study of dynamic-on (diffusion) and dynamic-off (structure) networks \cite{antoniades2015co,farajtabar2017coevolve}. The diffusion process affects the evolution of a network, and the change in the network impacts the life cycle of diffusion processes as well. Therefore, dynamics on and off affect each other \cite{weng2013role}. Paper \cite{7050284} proposes a time-delayed model for new link formation based on pre-existing links over time. Another paper finds new links by classical link prediction methods, then applies diffusion processes to new networks and evaluates and analyzes evolved networks' structural and diffusion processes \cite{vega2019evaluating}.

Table~\eqref{tab:relatedcompare} categorized different views of data missing occurrence in structure and diffusion class. Paying attention to each of them leads to solving different social network problems:

\begin{itemize}
	
	\item \textbf{Complete diffusion with partially structure network}\\
When a network is incomplete, it means some nodes and links are missing. As a result, forecasting and optimizing the diffusion process based on this partially observed network is not the same in reality. Estimating the cascade's influence on the unobserved part of the network is necessary. Seed selection for influence maximization is an application of this problem \cite{Phantom2016,sumith2018influence}.

	\item \textbf{Complete diffusion with no structure network}\\
	Network inference problems fall into this category. When the interaction topology of a network is entirely unreachable, an inference approach is put forward using visible independent measurements \cite{newman2018network}. An observational measurement could be the signals received from nodes or the cascades over the network \cite{gomez2012inferring,gomez2013modeling,rodriguez2014uncovering,ramezani2017dani,ji2020network,huang2019learning,gomez2013structure,tahani2016inferring,woo2020iterative,PhysRevE.96.012319,PhysRevE.98.062321,kefato2019refine}.
	
	\item \textbf{Partially diffusion with complete structure network}\\
	In the case of a complete graph of the network but incomplete information propagation, we have different diffusion problems, which include: \\
	(1) Identifying the future infected nodes in a cascade sequence by observing the incomplete and primary parts of a cascade sequence \cite{Topolstm2017,8826231,wang2017cascade,DeepDiffuse2018,MultiScale2019,wang2022casseqgcn,najar2012predicting,sun2022ms},\\
	(2) Discovering the source of diffusion \cite{farajtabar2015back,shi2022source},\\
	(3) Detecting missing parts of the cascade, such as nodes that were missed or infection times that were not observed \cite{sundareisan2015hidden},\\
	(4) Estimating the properties of complete cascades based on partially sampled diffusion \cite{CorrectingLeskovec2011}.
	
	\item \textbf{Partially diffusion with no structure network}\\
	Based on the few first windows of the cascade, some works attempt to predict the future size of the cascade without knowing how the nodes interact \cite{chen2022multi} or ranking the next infected nodes \cite{chen2022multi,yuan2021dyhgcn,wang2022cascade,wang2021dydiff}. The researchers also try to model cascade propagation.
	
	\item \textbf{No diffusion data with complete structure network}\\
	Modeling the pattern and path of a cascade over a network is the purpose of this category \cite{saito2008prediction}.

	\item \textbf{No diffusion data with partially structure network}\\
	The purpose of this category is to recover missing parts of a network. The omitted part can include nodes, and links \cite{kronecker2011}. There are numerous proposed models to predict lost links in a network when all nodes and some links are present \cite{
		kronecker2011,7752214,7403556,NIPS2008_8613985e,MFLink2018,
		mutlu2019review,menon2011link,li2016exploiting,jia2015learning,
		wai2022community,
		tang2023joint,chai2022network}.
	
	\item \textbf{Partially diffusion with partially structure network}\\
	According to our knowledge, there has been no prior research that specifically addresses the goals of this category. This paper aims to provide a method for simultaneous inference from cascade sequences and graph links when both have missing data.
	
\end{itemize}

\begin{table}
	\centering
	\caption{Classification of studies on diffusion process and network structure dimensions from the perspective of Complete (C) and Partial (P) data. (N) stands for Not considering that dimension.}
	\label{tab:relatedcompare}
	\smallskip\noindent
	\begin{tabular}{|l|c|c|c|c|} 
		\hline
		\multicolumn{1}{|c}{}               &                             & \multicolumn{3}{c|}{\textbf{Structure}}                                                                                              \\ 
		\cline{3-5}
		\multicolumn{1}{|c}{}               &                             & \textbf{C}                          & \textbf{P}                                             & \textbf{N}                            \\ 
		\hline
		\multicolumn{1}{|c|}{}              & \multirow{2}{*}{\textbf{C}} & \multirow{2}{*}{--}                 & \multirow{2}{*}{diffusion estimation}                    & \multirow{2}{*}{structure inference}  \\
		\multirow{5}{*}{\textbf{\rotatebox[origin=c]{90}{\textbf{Diffusion}}}} &                             &                                     &                                                        &                                       \\ 
		\cline{2-5}
		& \multirow{3}{*}{\textbf{P}} & cascade prediction                  & \multirow{3}{*}{\textbf{{focus of this paper}}} & \multirow{3}{*}{cascade prediction}   \\
		&                             & source detection                   &                                                        &                                       \\ 
		&                             & missing detection                   &                                                        &                                       \\ 
		\cline{2-5}
		& \multirow{2}{*}{\textbf{N}} & \multirow{2}{*}{diffusion modeling} & link prediction                                        & \multirow{2}{*}{--}                   \\
		&                             &                                     & network completion                                     &                                       \\
		\hline
	\end{tabular}
\end{table}

Currently, there is no previous work that is precisely related to the goals of the proposed method. It is worth noting that the works without considering partial observations, such as graph embedding or networks evolving over time, are not properly grounded in the literature of this paper. Also, one of the advantages of this paper is correcting the omitted data of diffusion against inferring the missing links, which are outside the scope of those works.

\section{Related Work}\label{sec:related work}
In this section, we summarize the related works with the assumption of incomplete data, either in structure or diffusion data, into three categories: "link prediction," "network inference," and "cascade correction and prediction."

\textbf{Link prediction}. Link prediction and network completion are two famous problems for solving missing data in structural networks. KronEM predicts missing nodes and links by combining an Expectation-Maximization (EM) approach with the Kronecker model of graphs \cite{kronecker2011}. Some works utilize side information such as node attribute \cite{7752214} or pairwise similarity between nodes \cite{7403556} to complete the network.\\
All nodes are visible in the link prediction works, while some links are omitted. There are numerous link prediction methods, including traditional supervised and unsupervised learning methods, probabilistic stochastic block model {\cite{NIPS2008_8613985e}}, matrix or tensor factorization \cite{MFLink2018}, locally-based algorithms, and deep learning based approaches \cite{mutlu2019review}. Although many link prediction methods exist, we only refer to those utilizing diffusion information to obtain additional information or matrix factorization to make their prediction.

\cite{menon2011link} utilizes a link prediction matrix factorization with features of nodes to predict unobserved links. \cite{li2016exploiting} employs diffusion features such as interactive activities of nodes in cascades against topology features for link prediction. Link prediction in \cite{jia2015learning} is modeled with matrix factorization using the similarity of retweeting information between pairs of users. In addition to predicting a link, paying close attention to the partially observed graph is critical for recognizing the community involved \cite{wai2022community}.

In LCPA \cite{chai2022network}, a maximum likelihood method was proposed to estimate structure networks by perturbing the adjacency matrix iteratively and correcting it. A partial binary structure network adjacency matrix with complete binary attribute information of nodes is used in JWNMF \cite{tang2023joint}. It proposes a non-negative matrix factorization method using two binary adjacency and attribute matrices sharing a topological hidden factor matrix.

\textbf{Network inference}. The goal is to infer the links of the underlying network using a set of cascades. For this purpose, Netrate \cite{rodriguez2014uncovering} solves a convex optimization, DANI \cite{ramezani2017dani} employs maximum likelihood while preserving the structural features of the network, and REFINE \cite{kefato2019refine} utilizes neural network and T-SVD for feature selection from nodes participating in different cascades to infer links between nodes that have similar embedding. Other attempts have been made to network inference, from static assumption requiring time stamp \cite{gomez2012inferring,gomez2013modeling,rodriguez2014uncovering,ji2020network,ramezani2017dani,kefato2019refine} or without infection time \cite{huang2019learning} to dynamic inference \cite{gomez2013structure,tahani2016inferring}. \cite{woo2020iterative} uses a base graph and complete cascade for estimating the graph path of diffusion iteratively. As a step towards using prior knowledge, some works employ in-degree distribution for nodes \cite{PhysRevE.96.012319} and measurements such as pathways, network properties, and information about the links or nodes \cite{PhysRevE.98.062321}.  

\textbf{Cascade correction and prediction}. If the initial and incomplete parts of the cascade are observed, how can the future infected nodes be identified? In specific, the LSTM architecture is used for predicting the next node in a cascade with the help of a complete network structure \cite{Topolstm2017}, estimating the next node \cite{8826231,MultiScale2019,sun2022ms} or finding the next infection time with the ranking of nodes as the next infection step \cite{wang2017cascade,DeepDiffuse2018}. If a propagation tree is available, \cite{kim2022models} uses the partial subgraphs to learn the representation of the full graph. NetFill finds the missing infected nodes and source of diffusion by observing incomplete and noisy cascades by proposing a model based on a Minimum Description Length (MDL) \cite{sundareisan2015hidden}.

To the best of our knowledge, none of the existing research incorporates simultaneous joint structure and diffusion data to solve the missing data problem. However, they attempt to recover the missing parts of both separately. This paper proposes the DiffStru method, which will be discussed in more detail in the following sections. DiffStru is more accurate than previous link prediction works because it uses structure and diffusion information to detect missing links simultaneously. Furthermore, DiffStru can solve more tasks than link prediction due to its ability to learn latent factors. If missing data occur in cascades, the network inference methods will perform less accurately than DiffStru. Moreover, DiffStru can infer the missing information of diffusion during the cascade sequence, while diffusion prediction methods are only used for predicting future infections. Meanwhile, DiffStru can be improved by decreasing its sensitivity to network density and considering the dynamic nature of the network structure.

\section{Modeling Information Diffusion and Network Structure}\label{sec:problem statement}

This section first introduces the notations and definitions used in the paper. Then, we state the joint inference problem of diffusion networks and structure from partially observed data.

We identify matrices, vectors, and scalars by uppercase bold-faced ($\mathbf{X} \in \mathbb{R}^{p \times q}$), lowercase bold-faced ($\mathbf{x} \in \mathbb{R}^{p \times 1}$) and normal lowercase ($x$) letters, respectively. All vectors are column vectors, and $\mathbf{x}_i$ is the $i$-th element of vector $\mathbf{x}$. {  Similarly}, $\mathbf{X}_{:j}$ is the $j$-th column, $\mathbf{X}_{i:}$ is the $i$-th row $\mathbf{X}$, and $\mathbf{X}_{ij}$ is the entry in the $i$-th row, and $j$-th column of $\mathbf{X}$. The $m \times m$ identity matrix is denoted by $\mathbf{I}_{m}$. $\mathbf{X}^T \in \mathbb{R}^{q \times p}$ returns the transpose of a matrix and $vec(\mathbf{X}) \in \mathbb{R}^{pq \times 1}$ is the linear operator flattening all the columns of the matrix to a column vector. For matrices $\mathbf{X} \in \mathbb{R}^{p \times q}$ and $\mathbf{Y} \in \mathbb{R}^{r \times s}$, $(\mathbf{X} \otimes \mathbf{Y}) \in \mathbb{R}^{pr \times qs}$ is referred to the Kronecker product of two matrices. {If $p=r$ and $q=s$}, then $(\mathbf{X} \circ \mathbf{Y}) \in \mathbb{R}^{p \times q}$ denotes the Hadamard product (elementwise). Table \eqref{tab:Hg} summarizes the notations we use in the proposed method.

We model a static social network with graph $\mathbb{G}=(V,E)$ where $V$ represents $N$ nodes (users of the network), $N=|V|$, and $E$ indicates the set of edges between nodes. Link $E_{ij}$ is formed from the relationship of node $i$ to node $j$. These edges represent directed and unweighted interactions (${E_{ij}} \ne {E_{ji}}$) without considering the self-links.
Suppose $\mathbf{G} \in \{0,1\}^{N \times N}$ is the corresponding asymmetric adjacency matrix of $\mathbb{G}$. Since the links of network graph $\mathbb{G}$ are not fully observable, the available structure of a network {  is a sub-graph $\mathbb{\acute{G}}=(V,\acute{E}) \subseteq \mathbb{G}$.} Let $\mathbf{\acute{G}}$ be a $N \times N$ binary adjacency matrix where the set of one value entries $\Omega^{+}=\{(i,j):\mathbf{\acute{G}}_{ij}=1\}$ denotes {existing} observed links of the network. 
$\Omega^{-}$
represents the missing links, while it may not surely mean to imply that there is no link. For each entry $(i,j) \in \Omega^{-}$, we may have $\mathbf{G}_{ij}=1$ or $\mathbf{G}_{ij}=0$ in the oracle network. 

In addition to the network structure, we have a set of information diffusion cascades $\mathbb{C}$ among the users represented by a matrix $\mathbf{C} \in \mathbb{R}^{N \times M}$. By denoting the overall spread of each information as cascade $c_{j}$; the element $\mathbf{C}(v_{i},c_{j})=t_{ij}$ represents that cascade $c_{j}$ reached the user $v_{i}$ and infected it at hit time $t_{ij} \in [0,T]  \cup {\infty}$, where $\infty$ used for users that are not infected by $c_{j}$ during the observation window $[0,T]$. We also assume that hit time is set to zero at the beginning of each cascade, and the cascade cannot infect each node more than once during its lifetime. A piece of information $j$ propagates over the structure network by transmitting through the links from an infected node to an uninfected node. Each cascade can be expressed as a $N \times 1$ vector $(\mathbf{{C}}_{:j})$. Since a contagion does not reach all $N$ observed network users, we {are faced with} a sparse vector for each cascade. Let $\mathbf{\acute{C}_{:j}}$ be the observed vector for cascade $j$, which contains only a subset of the infected users with their hit times. 
{Formally, $\Gamma^{+}_{j}=\{(i):\mathbf{C}_{ij}\in \mathbb{R}_{+} \wedge \mathbf{\acute{C}}_{ij}=\mathbf{C}_{ij} \}$} is a set  of  indices  of  the  observed  entries. The state of other nodes of the network $\Gamma^{-}_{j}=\{V \setminus \Gamma^{+}_{j}\}$ is hidden from us, while in the ground truth of the diffusion process, they may be infected or legitimate uninfected nodes. By aggregating all the observed vectors of $M$ different cascades in a matrix $(\mathbf{\acute{C}})$, the mask matrix for partially observable knowledge of diffusion can be represented with $\mathbf{\Gamma} \in \{0,1\}^{N \times M}$ which is a mapping from the collection of $M$ vector indices ($\Gamma^{+}_{j=1,...,M}$ and $\Gamma^{-}_{j=1,...,M}$) to a matrix space.\\
\begin{table}[t]
	\centering
	
	\caption{Major notations and symbols used for {DiffStru}}
	\label{tab:Hg}
	\begin{tabular}{ll}
		\toprule
		Symbol        & Description        \\
		\midrule
		$N$,$M$ & number of users, cascades  \\
		{$\mathbb{G}$,$V$,$E$,$\mathbf{G}$}&{network graph, set of nodes, set of edges (links), adjacency matrix}\\
		$\mathbf{C}$,$\mathbf{G}$ & $N \times M$ {original} diffusion matrix, $N \times N$ {original} structure matrix\\
		{$\mathbf{\Gamma}$}&{binary ${N \times M}$ mask matrix for partially observable knowledge of diffusion}\\
		{$\mathbf{\Omega}$}&{binary ${N \times N}$ mask matrix for the partially observed structure of the network}\\
		$\mathbf{\acute{C}}$,$\mathbf{\acute{G}}$ & observed diffusion matrix, observed structure matrix\\
		$D$   & dimension of latent low-rank factorize matrices \\
		$f(.)$ & logistic sigmoid function\\
		$\mathbf{\Xi}$ & $N \times N$ binary latent auxiliary variable for structure graph\\
		{$\mathbf{R}$}&{$N \times N$ stochastic auxiliary variable} \\
		{${\mathbf{W}_Y}$, $\mathbf{W}_U$, $\mathbf{W}_X$}&{covariance $M \times M$,$N \times N$,$N \times N$ matrices between pair of cascades, nodes}\\
		{ $\mu_\mathbf{\Xi}$}&{ conjugate Beta prior for observing pair of nodes}\\
		{ $\alpha_{1}$,$\alpha_{2}$}&{ hyper-parameters of a conjugate Beta prior assigned to the parameter $\mu_\mathbf{\Xi}$}\\
		{ $\sigma_{C}$,$\sigma_{R}$}&{ variance parameters for Gaussian distribution of $\mathbf{C}$,$\mathbf{R}$}\\
		{ $\delta_0$}&{ dirac delta distribution at zero}\\
		{ $\mathbf{P}$}&{ $N \times M$ infection probability of the nodes in cascades matrix}\\
		{ $\mathbf{A}$}&{  $N \times N$ infection transfer matrix between each nodes}\\
		$\mathbf{\Pi}$ & $N \times M$ binary hyper-parameter for diffusion information\\
		{$\mathbf{X},\mathbf{Y},\mathbf{U}$} & {$D \times N$ User, $D \times M$ Cascade, $D \times N$ Factor latent features}\\
		{ $\mathcal{N},\mathcal{RN}$}& { Normal, Rectified Normal distribution}\\
		{ ${\rm Ber},{\rm Beta},\mathcal{PG}$} & { Bernoulli, Beta, Polya-gamma distribution}\\
		$\mathcal{N}_{t}$ & Multivariate distribution for $t$ dimension vector\\
		$\mathcal{MN}_{tq}$ & $t \times q$ Matrix Gaussian distribution\\
		{$\otimes,\circ$ }&{kronecker product, Hadamard product (elementwise) }\\
		{$vec$} & {linear operator flattening all the columns of the matrix to vector}\\
		\bottomrule
	\end{tabular}
\end{table}
{ 
The problem we want to solve is:

\textbf{Given:} The partially observed network structure matrix $\mathbf{\acute{G}}$ and information diffusion matrix $\mathbf{\acute{C}}$.

\textbf{Goal:} Recover the non-observed links of a network by estimating matrix $\mathbf{\hat{G}}$ and obtain the approximated hit time of unobserved users of the network who are interacting hiddenly in diffusion processes with discovering matrix $\mathbf{\hat{C}}$.

\textbf{Our Approach:} Since our goal is to recover the hidden entries of two partially observed matrices, it can be modeled like a matrix completion problem. Matrix factorization is a prevalent and effective technique for matrix completion problems by approximating a given matrix $\mathbf{S} \in \mathbb{R}^{m \times n}$ as a product of two low-rank latent factor matrices $\mathbf{Q} \in \mathbb{R}^{m \times r}$ and $\mathbf{H} \in \mathbb{R}^{r \times n}$ with a constraint on $r \ll \min{(m,n)}$, such that $ \mathbf{S} \approx  QH$. The latent factors $Q$ and $H$ are $r$ Dimension representation matrices and can be interpreted as embedding for rows and columns of $\mathbf{S}$. These factors can be learned even if $\mathbf{S}$ is partially observed by minimizing the reconstruction error for the observed entries to recover the full $\mathbf{S}$ \cite{lee2001algorithms}.

We start from the idea that a node infection can signify how the nodes interact with each other in the network structure. However, the links between users can impact the diffusion process. As shown in Fig. \eqref{fig:trueobservered}, a network structure includes a set of users and social links between them while the information propagates among these users. A user would receive information if one of his friends (a user with whom he has a link to him) posted or reposted it. In conclusion, the diffusion process can reflect and drive the network's structure, demonstrating that two matrices $\mathbf{G}$ and $\mathbf{C}$ correlate. We will model this correlation property by sharing the same latent factors between $\mathbf{G}$ and $\mathbf{C}$. Therefore, the unobserved entries of $\mathbf{\acute{G}}$ and $\mathbf{\acute{C}}$ matrices can be estimated using the coupled matrix factorization method.}
\begin{figure}[ht]
	\noindent
	\centering
	\scalebox{0.55}{
		\input{./images/properties.tex}
	\begin{tikzpicture}[scale=0.5]
	\node[person,shirt=purple,minimum size=1.5cm,font=\footnotesize] (nD) at (17,15) {D};
	\node[alice,minimum size=1.5cm,font=\footnotesize] (nB) at (28,27) {B};
	\node[bob,minimum size=1.5cm,font=\footnotesize] (nA) at (0,25) {A};
	\node[businessman,minimum size=1.5cm,font=\footnotesize] (nE) at (-1,18) {E};
	\node[graduate,minimum size=1.5cm,font=\footnotesize] (nF) at (31,16) {F};
	\node[person,female,shirt=blue,minimum size=1.5cm,font=\footnotesize] (nC) at (15,25) {C};

	\draw  [->, ultra thick, \arrowColor] (nA.east) to (nC.west);
	\draw  [->, ultra thick, \arrowColor] ($(nC.south)+(0,-0.6)$) to ($(nD.north)+(-0.4,0.1)$);
	\draw  [->, ultra thick, \arrowColor] (nD.south west) to (nE.south east);
	\draw  [->, ultra thick, \arrowColor] ($(nD.east)+(0,-1)$) to ($(nF.west)+(0,-1)$);
	\draw  [->, ultra thick, \arrowColor] (nB.south west) to ($(nD.north)+(0.1,0)$);
	\draw  [->, ultra thick, \arrowColor] (nC.south west) to (nE.east);
	
	\draw  [<-, ultra thick, \arrowColor] ($(nC.south west) + (0,-0.4)$) to ($(nE.east) + (0, -0.4)$);
	\draw  [<-, ultra thick, \arrowColor] ($(nB.south west) + (0,0.4)$) to ($(nD.north) + (0,0.4)$);
	\draw  [<-, ultra thick, \arrowColor] ($(nA.east) + (0, -0.4)$) to ($(nC.west) + (0, -0.4)$);

	\node[ellipse callout,draw,inner sep=2pt,fill=\messageOneColor,align=center,callout absolute pointer=(nA.mouth),above right= 5pt and 0pt of nA.north east,font=\tiny] at (nA.north){	
		- - - - - - - - -  \\
		- - - - - - - - -  \\
		08:50 AM 23 Nov 2022};
	
	\node[ellipse callout,draw,inner sep=2pt,fill=\messageOneColor,align=center,callout absolute pointer=(nC.mouth)),above right= 5pt and 0pt of nC.north east,font=\tiny] at (nC.north){	
	- - - - - - - - -  \\
	- - - - - - - - -  \\
	01:35 PM 23 Nov 2022};

	
	\node[ellipse callout,draw,inner sep=2pt,fill=\messageOneColor,align=center,callout absolute pointer=(nE.west)),above left= 1pt and 0pt of nE.north west,font=\tiny] at (nE.north west){	
	- - - - - - - - -  \\
	- - - - - - - - -  \\
	08:25 PM 23 Nov 2022};

	
	\node[ellipse callout,draw,inner sep=2pt,fill=\messageTwoColor,align=center,callout absolute pointer=(nD.mouth)),above right= 5pt and 0pt of nD.east,font=\tiny] at ($(nD.east)+(1.5,0)$){	
		- - - - - - - - -  \\
		- - - - - - - - -  \\
		03:30 PM 23 Nov 2022};
	
	
	\node[ellipse callout,draw,inner sep=2pt,fill=\messageTwoColor,align=center,callout absolute pointer=(nF.mouth)),above right= 5pt and 0pt of nF.north west,font=\tiny] at (nF.north){	
		- - - - - - - - -  \\
		- - - - - - - - -  \\
		01:35 PM 23 Nov 2022};

	
	\node[ellipse callout,draw,inner sep=2pt,fill=\messageTwoColor,align=center,callout absolute pointer=(nB.west)),above left= 1pt and 0pt of nB.north west,font=\tiny] at (nB.north west){	
		- - - - - - - - -  \\
		- - - - - - - - -  \\
		09:45 AM 24 Nov 2022};

\end{tikzpicture}
	}
	\caption{
		{  This illustrates the relationship between partially observed diffusion information and network structure. A link from "$D$" to "$E$" means $D$ is following $E$, so $E$ is the followee, and $ D$, the follower monitor his post.}
	}
	\label{fig:trueobservered}
\end{figure}
\\

\section{Proposed Method}\label{sec:proposed solution}
This section presents the proposed model framework and how to infer the latent factor matrices. 

\subsection{Model Framework}
{ The structural network $\mathbf{G}$ is a Bernoulli distribution with a stochastic auxiliary variable $\mathbf{R}_{ij}$ in the latent feature space:}
\begin{equation}
\small
\label{eq:MF_G_simple}
\begin{gathered}
\mathbf{G}_{ij}={\rm{Ber}}(f(\mathbf{R}_{ij}))\;\;\;\\
\mathbf{R}_{ij}=\mathbf{X}_{:i}^{T}\mathbf{U}_{:j}+\varepsilon_{ij} \;\;\; \varepsilon_{ij} \sim \mathcal{N}(0,\sigma_{R}^2) \\
\end{gathered}       
\end{equation}
where $f(t)=\cfrac{1}{1+e^{-t}}$ is the logistic sigmoid function. {  As shown in Fig. \eqref{fig:logisticLink}, $\mathbf{R}_{ij}$ is constructed using Gaussian distribution with low-rank matrices $\mathbf{X} \in \mathbb{R}^{D \times N}_{+}$ and $\mathbf{U} \in \mathbb{R}^{D \times N}$ that represent user-specific and factor specific latent features, respectively. In order to limit the range of $\mathbf{R}_{ij}$ to $[0,1]$, a logistic sigmoid function $f(.)$ is used. The Bernoulli distribution, which has$0$ and $1$ samples., is applied}
%
%
%
%
%
\begin{figure}[ht]
	\centering
	\noindent
	\scalebox{.70}{
		\input{./images/properties.tex}
\begin{tikzpicture}
	\begin{scope}[scale=0.8]
		\draw[thin] (-2, -0.05) -- (2, -0.05);
		\draw[thin] (0, -0.1) -- (0,2);
		\node [font=\scriptsize] at (0,  2.5) {$\mathcal{N}(X_{:i}^{T}U_{:j}, \sigma_{R}^{2})$};
		\draw[color=plot, thick, samples=100, domain=-4:4, yscale=5, xscale=0.5] plot (\x,{(1/sqrt(2*pi))*exp(-0.5*(\x*\x))});
	\end{scope}
	
	\begin{scope} [xshift=1cm]
		\draw[>-stealth] (0, 0.8) -- (0.3, 0.8) node[above] {$\sim$} -- (0.6, 0.8) ;
		\node [right] at (0.6, 0.8)  {$R_{ij}$};
		\draw[>-stealth] (1.3, 0.8) -- (1.9, 0.8);
		
		\begin{scope} [xshift=3.1cm, scale=0.8]
			\draw[thin] (-2, -0.05) -- (2, -0.05);
			\draw[thin] (0, -0.01) -- (0,2);
			\draw[color=plot, thick, samples=100, domain=-4:4, yscale=2, xscale=0.5] plot (\x,{1/(1 + exp(-\x))});
			\node [left] at (0, 1) {$0.5$};
		\end{scope}
	\end{scope}
	
	\begin{scope}[xshift=6.2cm] 
		\draw[>-stealth] (-1.6, 0.8) -- (-1, 0.8);
		\node [right] at (-0.9, 0.8) {$q=f(R_{ij})$};
		\draw [>-stealth] (0.9, 0.8) -- (1.5, 0.8);
		
		\fill[bar1] (2.4,0) rectangle (2,0.5);
		\node [above] at (2.2, 0.5) {$q$};
		\fill[bar2] (3,0) rectangle (2.6,1.3);
		\node [align=left, above] at (2.8, 1.3) {$1-q$};
				
		\draw [thin] (2, 0) -- (3, 0);
		
		\draw [>-stealth] (3.5, 0.8) -- (3.8, 0.8) node[above]{$\sim$} --
		(4.1, 0.8);
		
		\node [right] at (4.5, 0.8) {$G_{ij}$};
		
	\end{scope}
\end{tikzpicture}
	}
	\caption
	{
		{Mathematical} Diagram of the generalized linear model for estimating the underlying link existence ($\sim$ means sampling from the left distribution).
	}
	\label{fig:logisticLink}
\end{figure}

{  A binary latent auxiliary variable $\mathbf{\Xi}$ is employed as a link observer variable to express $\mathbf{{G}}$, which takes binary values.
\begin{itemize}
	\item  $\mathbf{\Xi}_{ij}=0$ means that we did not observe $\mathbf{G}_{ij}$ (link existence between nodes $v_i$ and $v_j$ has not been investigated), so $\mathbf{\acute{G}}_{ij}=0$. 
	\item $\mathbf{\Xi}_{ij}=1$ implies that $\mathbf{{G}}_{ij}$ was observed and based on $(i,j)$ included in ${E}_{ij}$ ($\mathbf{G}_{ij}=1$) or not $\mathbf{G}_{ij}=0$, $\mathbf{{\acute{G}}}_{ij}$ takes one or zero values, respectively.
\end{itemize}
} While $\mathbf{{{G}}}_{ij}$ is conditional on the value of $\mathbf{\Xi}_{ij}$, the distribution function of $\mathbf{G}_{ij}$ can be modeled as a mixture of two distributions with mixing weight $\mathbf{\Xi}_{ij}$. The two components of $\mathbf{{G}}_{ij}$ are assumed to be a Bernoulli distribution with a probability of success $f(\mathbf{R}_{ij})$ and a Dirac delta distribution at zero:
\begin{equation}
\small
\label{eq:G_Prob}
\begin{gathered}
\mathbf{{G}}_{ij} \sim \mathbf{\Xi}_{ij}{\rm Ber}(f({R}_{ij}))+(1-\mathbf{\Xi}_{ij})\delta_0
\end{gathered}    
\end{equation}

We consider the Bernoulli distribution for latent auxiliary variable $\mathbf{\Xi}_{ij} \sim {\rm Ber}(\mu_\mathbf{\Xi})$. By assuming equal probability for observing all pairs of nodes {$(v_i,v_j)$ that ${{i,j \in V},{i \ne j}}$}
, a conjugate Beta prior with hyper-parameters $\alpha_1$ and $\alpha_2$ are assigned to the parameter $\mu_\mathbf{\Xi} \sim {\rm Beta}(\alpha_1,\alpha_2)$.

The cascade matrix $\mathbf{{C}}$ can be modeled as:
\begin{equation}
\small
\label{eq:MF_C_simple}
\mathbf{C}_{ij}=\mathbf{X}_{:i}^{T}\mathbf{Y}_{:j}+\eta_{ij} \;\;\;\;\;\; \eta_{ij} \sim \mathcal{N}(0,\sigma_{C}^2)        
\end{equation}
{where $\mathbf{X} \in \mathbb{R}^{D \times N}_{+}$ and $\mathbf{Y} \in \mathbb{R}^{D \times M}_{+}$} are low-rank matrices representing the user and cascade latent features with nonnegative entries, and $\eta_{ij}$ denotes the residual noise sampled independently from zero-mean Gaussian distribution with variance $\sigma_{C}^2$. 
The stochastic modeling of the hyper-parameter matrix $\mathbf{\Pi}$ as an observer for $\mathbf{{C}}$ leads to an impossible inference model. Therefore, we resort to a deterministic model. The hyper-parameter $\mathbf{\Pi}_{ir}$ indicates that we have observer knowledge about the infection of node $v_i$ in cascade $c_r$ or not.
Fig. \eqref{fig:PGM} shows the generative Bayesian probabilistic representation of the proposed method for jointly inferring the Diffusion and Structure of the network called "DiffStru." 

Our basic idea for joint inference is to incorporate the shared latent factor $\mathbf{X}$ between network structure and diffusion matrices and choose the suitable prior distributions for considering the side information.

When a user account is private, no information about his diffusion behavior is available; hence, the corresponding rows of latent factors will be empty. However, we can handle these empty rows of matrices by integrating the side information as prior knowledge for capturing the correlation between users or cascades. {Here, a zero-mean multivariate Gaussian distribution is employed as a conjugate prior of $U$ as $\mathbf{U}_{d:} \sim \mathcal{N}_{N}\left(0,{\mathbf{W}_U} \right)$ and $X$ and $Y$ are drawn from a rectified multivariate Gaussian prior with zero mean: $\mathbf{X}_{d:} \sim \mathcal{RN}_{N}\left(0,{\mathbf{W}_X} \right)$ and $\mathbf{Y}_{d:} \sim \mathcal{RN}_{M}\left(0,{\mathbf{W}_Y} \right)$ for supporting nonnegative constraints.}
We exploit the diffusion and topological metrics as prior distributions for each row of the latent matrices $\mathbf{Y}_{d:}$, $\mathbf{U}_{d:}$, and $\mathbf{X}_{d:}$, with covariance matrices ${\mathbf{W}_Y}$, $\mathbf{W}_U$, and $\mathbf{W}_X$, respectively. Each element of full covariance matrices ${\mathbf{W}_Y \in \mathbb{R}^{M \times M}}$ (${\mathbf{W}_U \in \mathbb{R}^{N \times N}}$ or ${\mathbf{W}_X \in \mathbb{R}^{N \times N}}$) in these distributions capture the relationship between the pair of cascades (users) and also the correlation between different features of a cascade (user). These covariance matrices will apply the covariances between rows and columns of matrices in the priors \cite{zhou2012kernelized}. We will show later in Section \eqref{sec:hypersetting} how to initialize these covariances. The distributions are expressed as Algorithm \eqref{algo:generativeprocess}.

\begin{algorithm}
	\caption{Generative process for DiffStru graphical model}
	\label{algo:generativeprocess}
	\SetAlgoLined
	$\mu_\mathbf{\Xi} \sim {\rm Beta}(\alpha_1,\alpha_2)$\\
	$\mathbf{Y}_{d:} \mathop \sim \limits^{iid} \mathcal{RN}_{M}\left(0,{\mathbf{W}_Y} \right), \forall d=1,...,D$\\
	$\mathbf{X}_{d:} \mathop \sim \limits^{iid} \mathcal{RN}_{N}\left(0,{\mathbf{W}_X} \right), \forall d=1,...,D$\\
	$\mathbf{U}_{d:} \mathop \sim \limits^{iid} \mathcal{N}_{N}\left(0,{\mathbf{W}_U} \right), \forall d=1,...,D$\\
	\ForEach {user $i=1,...,N$}
	{
		
		\ForEach {user $j=1,...,N$}
		{
			$\mathbf{\Xi}_{ij} \mathop \sim \limits^{iid} {\rm Ber}(\mu_\mathbf{\Xi})$\\
			$\mathbf{R}_{ij}  \mathop \sim \limits^{iid} \mathcal{N}(\mathbf{X}_{:i}^{T}\mathbf{U}_{:j},\sigma_{R}^2)$\\
			$\mathbf{{G}}_{ij} \mathop \sim \limits^{iid} \mathbf{\Xi}_{ij}({\rm Ber}(f({R}_{ij})))+(1-\mathbf{\Xi}_{ij})(\delta_0)$
		}
		
		\ForEach {cascade $r=1,...,M$}
		{
			$\mathbf{C}_{ir}  \mathop \sim \limits^{iid} 
			\mathbf{\Pi}_{ir}(\mathcal{N}(\mathbf{X}_{:i}^{T}\mathbf{Y}_{:r},\sigma_{C}^2))
			$
		}
	}
\end{algorithm}

In the general formulation based on the graphical model in Fig. \eqref{fig:PGM}, the joint distribution over the observed, latent, and auxiliary random variables given the hyper-parameters\\ $\Theta=\{\mathbf{W}_Y,\mathbf{W}_X,\mathbf{W}_U,\alpha_1,\alpha_2,\sigma_{C}^2,\sigma_{R}^2,\mathbf{\Pi}\}$ is given by: 
\begin{equation}
\small
\begin{gathered}
\label{eq:joint}
P(\mathbf{X},\mathbf{U},\mathbf{Y},\mathbf{R},\mathbf{\Xi},\mu_{\Xi}, \mathbf{{G}},\mathbf{{C}}|\Theta) =  
\mathop{\mathlarger{\prod}} \limits_{i=1}^{N} \mathop{\mathlarger{\prod}} \limits_{j=1}^{M}\big(
\mathcal{N}(
\mathbf{R}_{ij}|
\mathbf{X}_{:i}^{T}\mathbf{U}_{:j},\sigma_{R}^2)
\times
[
\mathcal{N}(
\mathbf{C}_{ij}|
\mathbf{X}_{:i}^{T}\mathbf{Y}_{:j},\sigma_{C}^2)
]^{\mathbf{\Pi_{ij}}}\\
\times 
[
{\rm Ber}(\mathbf{G}|f(\mathbf{R}_{ij}))]^{\mathbf{\Xi}_{ij}}
\times  
{\rm Ber}(\mathbf{\Xi}_{ij}|\mu_\mathbf{\Xi})
{\rm Beta}(\mu_\mathbf{\Xi}|\alpha_1,\alpha_2)\big)
\times \mathop{\mathlarger{\prod}} \limits_{d=1}^{D} 
\big(
\mathcal{N}_{N}\left(\mathbf{U}_{d:}|0,{\mathbf{W}_U} \right)	\\
\times
\mathcal{RN}_{M}\left(\mathbf{Y}_{d:}|0,{\mathbf{W}_Y} \right)
\times
\mathcal{RN}_{N}\left(\mathbf{X}_{d:}|0,{\mathbf{W}_X} \right)
\big)
\end{gathered}
\end{equation}
{We should learn the posterior distribution to estimate the unobserved data using a Bayesian approach. 
	A posterior with an intractable integral will be obtained using Bayes's rule. For approximating these integrals, the MCMC algorithm is a common approximation approach that tries to obtain a sufficient number of samples of the target distribution for a dependable inference \cite{carlo2004markov}.}
\begin{figure}[t]
	\centering
	\scalebox{.6}{
%
%
%

\usetikzlibrary{calc}
\tikzset{
	latent/.style={draw, minimum width=8mm, shape=circle, ultra thick, black}
}
\tikzset{
	auxiliary/.style={draw,dashed, minimum width=8mm, shape=circle, ultra thick, black}
}
\begin{tikzpicture}

  \node[latent] (Rij) {$\mathbf{{R}}_{ij}$};
  \node[obs,right=of Rij, yshift=-0.1cm] (Gij) {$\mathbf{\acute{G}}_{ij}$};
  \node[below=of Rij, yshift=-1.8cm]  (sigmaR) {${\sigma_R}^2$};
  \node[latent, above=of Gij, xshift=0cm] (Eij) {$\mathbf{\Xi}_{ij}$};
  \node[latent, right=of Eij, xshift=1.8cm]  (miuXi) {$\mu_{\mathbf{\Xi}}$};
  \node[const, above=of miuXi, xshift=-0.3cm,yshift=0.2cm]  (alpha1) {${\alpha_1}$};
  \node[const, above=of miuXi, xshift=0.4cm,yshift=0.2cm]  (alpha2) {${\alpha_2}$};
  \node[latent, above=of Rij, xshift=-1.2cm, yshift=0.3cm]  (U) {$\mathbf{{U}}_{d:}$};
  \node[const, above=of U, yshift=0.3cm]  (WU) {$\mathbf{{W}}_{U}$};
  \node[latent, left=of Rij, xshift=-0.4cm, yshift=0.5cm] (Xi) {$\mathbf{{X}}_{d:}$};
  \node[obs, below=of Xi, xshift=-1.2cm] (Cij) {$\mathbf{\acute{C}}_{ir}$};
  \node[const, left =of Cij, yshift=-0.3cm] (Pi) {$\mathbf{\Pi}_{ir}$};
  \node[latent, left=of Xi, xshift=-2.7cm] (Yj) {$\mathbf{{Y}}_{d:}$};
  \node[const, above=of Xi,yshift=1.6cm]  (WX) {$\mathbf{{W}}_{X}$};
  \node[const, above=of Yj, yshift=0.4cm, xshift=-1cm]  (WY) {$\mathbf{{W}}_{Y}$};
  \node[const, below=of Cij,xshift=0cm,yshift=-0.4cm]  (sigmaC) {${\sigma_C}^2$};


  \edge {miuXi} {Eij};
  \edge {Xi,U,sigmaR} {Rij};
  \edge {Eij,Rij} {Gij};
  \edge {alpha1,alpha2} {miuXi};
  \edge {WU} {U}
  \edge {WX} {Xi}
  \edge {WY} {Yj}
  \edge {Xi,Yj,sigmaC,Pi} {Cij}

%
%
{\tikzset
	{
	plate caption/.append style={below=2.1cm of Gij.east,xshift=1cm}
	}
	\plate 
	[shape=rectangle,align=right, inner sep=0.41cm, yshift=-0.2cm, xshift=0.1cm,color=black]
	{p1} {(Rij)(Xi)(Cij)(Gij)(U)(Pi)} {$i,j=1,...,N$} ;
}

{\tikzset
	{
		plate caption/.append style={below=2.5cm of Yj.south west,xshift=0cm}
	}
	
	\plate 
	[shape=rectangle,align=right, inner sep=0.2cm, yshift=0cm, xshift=-0.1cm,color=black]
	{p2} {(Cij)(Yj)(Pi)} {$r=1,...,M$};
}

\plate {p3} {(Xi)(Yj)(U)} {$d=1,..,D$} ;

{\tikzset
	{
		plate caption/.append style={below=0.5cm of Gij.south}
	}

\plate 
[shape=rectangle,align=right, inner sep=0.2cm, yshift=0.15cm, xshift=0cm,color=black]
{p4} {(Gij)} {$tr(\Omega^T\Omega)$};
}

{\tikzset
	{
	plate caption/.append style={below=0.5cm of Cij.south}
}

\plate 
[shape=rectangle,align=right, inner sep=0.2cm, yshift=0.15cm, xshift=0cm,color=black]
	{p5} {(Cij)} {$tr(\Gamma^T\Gamma)$};
}


\end{tikzpicture}

	}
	\caption{
		The generative Bayesian probabilistic representation of the proposed model (DiffStru).
	}
	\label{fig:PGM}
\end{figure}
\subsection{Model Inference}
There are various ways to sample a distribution. In this case, we use the Gibbs method as an MCMC technique by iterative sampling one variable from a conditional distribution with fixing the remaining variables. In each iteration, all random variables are updated based on the previous value of others. We randomly initialize the values of random variables. To minimize the effect of random initialization, we should drop the first early samples, which is known as choosing the sufficient burn-in period. Also, thinning the chain is needed to avoid biased estimation of correlated samples and reduce processing and storing costs. Thinning is done by not considering the correlated samples and averaging over every $t$-th iteration \cite{carlo2004markov}. The details of the following equations can be found in "Proposed Method Details," Section 2 of the supplementary document for this paper \cite{supplementary}.\\
\textbf{Sampling} $\mathbf{Y}$: 
{Using the $vec$ operator, the posterior of $Y$ is a Gaussian distribution multiply unit step function, where the covariance $\mathbf{\Sigma_Y}$ and mean $\mathbf{\mu_Y}$ are given by:}
\begin{equation}
\small
\label{eq:Ycovariance}
\begin{gathered}
\mathbf{\Sigma_Y}=
[
(
[(\mathbf{I}_{M} \otimes \mathbf{X}) \circ
({\sigma_{C}^{-2}}\mathbf{1}_{MN} \otimes vec({\mathbf{\Pi}}))]
(\mathbf{I}_{M} \otimes \mathbf{X}^T)) 
+(
{\mathbf{I}_{D}} \otimes {\mathbf{W}_Y}^{-1}
)
]^{-1}\\
\mathbf{\mu}_{Y}={\Sigma_Y}([(
\mathbf{I}_{M} \otimes \mathbf{X})
\circ({\sigma_{C}^{-2}}\mathbf{1}_{MN} \otimes vec({\mathbf{\Pi}}))] vec({\mathbf{C}}))
\end{gathered}
\end{equation}
{When the covariance matrix is identity, for each cascade $j$, we can sample latent factor from a $D$-dimensional normal distribution multiply unit step function where the covariance matrix and the mean vector are given by:
	\begin{equation}
	\small
	\label{eq:Ycovariance2}
	\begin{gathered}
	\mathbf{\Sigma_Y}^{(j)} = \left[\sigma_{c}^{-2}(\mathbf{X}\circ\mathbf{\Pi}_{:j})\mathbf{X^T}+\mathbf{I}_D\right]^{-1} \\
	\mu_\mathbf{Y}^{(j)}=\sigma_{c}^{-2}\mathbf{\Sigma_Y}^{(j)}(\mathbf{X}\circ\mathbf{\Pi}_{:j})\mathbf{C}_{:j}
	\end{gathered}
	\end{equation}
}
\\ \textbf{Sampling} $\mathbf{\Xi}$: {As discussed in the previous section, $\mathbf{\Xi}_{ij}=1$ when $\mathbf{\Gamma}_{ij}=1$, and hence the conditional posterior density of $\mathbf{\Xi}$ is:}
\begin{equation}
\small
\centering
\begin{gathered}
\label{eq:SamplingEdetails}
P(\mathbf{\Xi}_{ij}|\mathbf{G}_{ {ij}}=0,-)
=
[\mathbf{\Xi}_{ij}(1-f(\mathbf{R}_{ij})
)
+
(1-\mathbf{\Xi}_{ij})\mathbb{I}(
\mathbf{G}_{ij}=0)]
\times {\mu_{\mathbf{\Xi}}}^{\mathbf{\Xi}_{ij}}{(1-\mu_{\mathbf{\Xi}})}^{1-\mathbf{\Xi}_{ij}}
\end{gathered}
\end{equation}
But this value should be inferred for elements when $\mathbf{\Gamma}_{ij}=0$. In this case $P(\mathbf{\Xi}_{ij})$ is a Bernoulli distribution:
\begin{equation}
\small
\begin{gathered}
\label{eq:BernoulliE}
P(\mathbf{\Xi}_{ij}=0|\mathbf{G}_{ {ij}}=0,-) \propto {(1-\mu_{\mathbf{\Xi}})}\\
P(\mathbf{\Xi}_{ij}=1|\mathbf{G}_{ {ij}}=0,-)\propto \mu_{\mathbf{\Xi}}(1-f(\mathbf{R}_{ij}))\\
\end{gathered}
\end{equation}
Therefore, the sampling posterior is:
\begin{equation}
\small
\begin{gathered}
\label{eq:FinalSamplingE}
P(\mathbf{\Xi}_{ij}=1|\mathbf{G}_{ {ij}}=1,-)=1\\
P(\mathbf{\Xi}_{ij}|\mathbf{G}_{ {ij}}=0,-)={\rm{Ber}}(\xi) \;\;\;\;\;\; 
\xi=\cfrac{\mu_{\mathbf{\Xi}}-\mu_{\mathbf{\Xi}}f(\mathbf{R}_{ij}) }{1-\mu_{\mathbf{\Xi}}f(\mathbf{R}_{ij})}
\end{gathered}
\end{equation}
%
\textbf{Sampling} $\mu_{\mathbf{\Xi}}$: Due to the use of conjugate prior for $\mu_{\mathbf{\Xi}}$, their conditional distribution will be a Beta distribution. Equation \eqref{eq:Samplingmu} shows the details for $\mu_{\mathbf{\Xi}}$:
\begin{equation}
\small
\centering
\begin{gathered}
\label{eq:Samplingmu}
P(\mathbf{\mu_{\Xi}}|-)
\sim \rm{Beta}(\mu_{\Xi}|\alpha_1+\mathop{\mathlarger{\sum}_{i,j}}\mathbf{\Xi}_{ij}
,
\alpha_2+N^2-\mathop{\mathlarger{\sum}_{i,j}}\mathbf{\Xi}_{ij}
)
\end{gathered}
\end{equation}
\textbf{Sampling} $\mathbf{U}$: {Given the other latent variables, the posterior of $\mathbf{U}$ is, we can sample { $vec(\mathbf{U})$} from the normal distribution, where the covariance $\Sigma_U$ and mean $\mu_U$ are given by:}
\begin{equation}
\small
\label{eq:Ucovariance}
\begin{gathered}
\mathbf{\Sigma_U}=
[(({\sigma_{R}^{-2}\mathbf{I}_{N}})
\otimes
({\mathbf{X}}\mathbf{X}^T))
+
(\mathbf{I}_{D} \otimes {\mathbf{W}_U}^{-1} )
]^{-1}\\
\mathbf{\mu}_{U}=\mathbf{\Sigma_U}[(\sigma_{R}^{-2}\mathbf{I}_{N} \otimes \mathbf{X})vec(\mathbf{R})]
\end{gathered}
\end{equation}
{With the identity covariance matrix, for each user $i$, we can sample latent factor from a $D$-dimensional normal distribution with the following mean and covariance.
	\begin{equation}
	\small
	\label{eq:Ucovariance2}
	\begin{gathered}
	\mathbf{\Sigma}_U = \left[\sigma_r^{-2}\mathbf{X}\mathbf{X}^T+I_{D}\right]^{-1} ;\;\
	\mu_U^{(i)} = \sigma_r^{-2}\mathbf{\Sigma_U}\mathbf{X}\mathbf{R}_{:i}
	\end{gathered}
	\end{equation}
}
There is a problem in sampling the remaining variables of the joint distribution \eqref{eq:joint} because of the logistic likelihood
function, which is not conjugate with other Gaussian terms. Hence, we utilize 
the Polya-Gamma latent variables \cite{polson2013bayesian}
by adding an auxiliary random variable $\mathbf{\Lambda}_{ij}$ (Fig. \eqref{fig:PolyaGamma}) for approximating the logistic likelihood with a Gaussian distribution that can easily multiply with prior normal distributions.

%
\begin{figure}[t]
	\centering
	\scalebox{0.55}{
%
%
%

\usetikzlibrary{calc}
\tikzset{
	latent/.style={draw, minimum width=8mm, shape=circle, ultra thick, black}
}
\tikzset{
	auxiliary/.style={draw,dashed, minimum width=8mm, shape=circle, ultra thick, black}
}
\begin{tikzpicture}

  \node[latent] (Rij) {$\mathbf{{R}}_{ij}$};
  \node[obs,right=of Rij, yshift=-0.1cm] (Gij) {$\mathbf{\acute{G}}_{ij}$};
  \node[latent, above=of Gij, xshift=0cm] (Eij) {$\mathbf{{E}}_{ij}$};
  \node[auxiliary, right=of Gij, xshift=0cm] (lambdaij) {$\Lambda_{ij}$};

    \edge {Eij,Rij,lambdaij} {Gij};

%
%
{\tikzset
	{
	plate caption/.append 
	}
	\plate 
	{p1} {(Rij)(lambdaij) (Eij)(Gij)} {$i,j=1,...,N$} ;
}

\end{tikzpicture}

	}
	\caption{
		Adding Polya-gamma auxiliary variables ($\mathbf{\Lambda}_{ij}$) for presenting the model as conditionally conjugate.
	}
	\label{fig:PolyaGamma}
\end{figure}
\begin{equation}
\small
\centering
\begin{gathered}
\label{eq:LogistictoNormal_v1}
P(\mathbf{G}_{ij}|\mathbf{R}_{ij},\mathbf{E}_{ij}=1)
=f(\mathbf{R}_{ij})
=\cfrac{1}{1+e^{(-\mathbf{R}_{ij})}}=
\cfrac{e^{(\mathbf{R}_{ij})}}{e^{(\mathbf{R}_{ij})}+1}\\
=\cfrac{1}{2}e^{(\mathbf{G}_{ij}-\cfrac{1}{2}){\mathbf{R}_{ij}}}\mathop{\mathlarger{\int}}_{0}^{\infty}{e^{\cfrac{-\mathbf{\Lambda}_{ij}{\mathbf{R}_{ij}}^2}{2}}P(\mathbf{\Lambda}_{ij})d{\mathbf{\Lambda}_{ij}}}
\end{gathered}
\end{equation}
where, $P(\mathbf{\Lambda}_{ij}) \sim PG(\mathbf{\Lambda}_{ij}|1,0)$. By conditioning \eqref{eq:LogistictoNormal_v1} on auxiliary variable, we obtain:
\begin{equation}
\small
\centering
\begin{gathered}
\label{eq:LogistictoNormal_v2}
P(\mathbf{G}_{ij}|\mathbf{R}_{ij},\mathbf{E}_{ij}=1,\mathbf{\Lambda}_{ij})\propto
exp((\mathbf{G}_{ij}-0.5)\mathbf{R}_{ij}-\cfrac{1}{2}\mathbf{\Lambda}_{ij}\mathbf{R}_{ij}^2)
\end{gathered}
\end{equation}
\textbf{Sampling} $\mathbf{\Lambda}$: The sampling of $\mathbf{\Lambda}$ is given by:
\begin{equation}
\small
\centering
\begin{gathered}
\label{eq:Samplinglambda}
P(\mathbf{\Lambda}_{ij}|-)\propto
P(\mathbf{G}_{ij}|\mathbf{R}_{ij},\mathbf{\Xi}_{ij}=1,\mathbf{\Lambda}_{ij})P(\mathbf{\Lambda}_{ij}) \sim \rm{PG}(\mathbf{\Lambda}_{ij}|1,\mathbf{R}_{ij})
\end{gathered}
\end{equation}
\textbf{Sampling} {$\mathbf{X}$: The posterior distribution of { ${vec(\mathbf{X}^{T})}$ is a normal distribution with following covariance and mean:}}
\begin{equation}
\small
\begin{gathered}
\label{eq:Xcovariance}
\mathbf{\Sigma_X}=
[
[(
(\mathbf{Y} \otimes \mathbf{I}_{N})
\circ ({\sigma_{C}^{-2}}(vec({\mathbf{\Pi}}))^{T}))
(\mathbf{Y}^T \otimes \mathbf{I}_{N})]
+
[\sigma_{R}^{-2}(\mathbf{U} \otimes \mathbf{I}_{N})
(\mathbf{U}^T \otimes \mathbf{I}_{N})]
+
({\mathbf{I}_{D}} \otimes {\mathbf{W}_X}^{-1})
]^{-1}\\
\mathbf{\mu}_{X}^T=						{\Sigma_X}[
[(
(\mathbf{Y} \otimes \mathbf{I}_{N})
\circ ({\sigma_{C}^{-2}}(vec({\mathbf{\Pi}}))^{T}))
vec(\mathbf{C})]+
(\sigma_{R}^{-2} (\mathbf{U} \otimes \mathbf{I}_{N})vec(\mathbf{R}))
]
\end{gathered}
\end{equation}
{ 
Assuming an initialized identity covariance, for each user $i$, $\mathbf{X}$ can be sampled from a $D$-dimensional normal distribution with following parameters:
}
	\begin{equation}
	\small
	\begin{gathered}
	\label{eq:Xcovariance2}
	\mathbf{\Sigma_X}^{(i)} = \left[\sigma_{c}^{-2}(\mathbf{Y}\circ\mathbf{\Pi}_{i:})\mathbf{Y}^T+\sigma_{r}^{-2}\mathbf{U}\mathbf{U}^T+\mathbf{I}_D\right]^{-1}\\
	\mu_\mathbf{X}^{(i)} = \mathbf{\Sigma_X}^{(i)}\left[\sigma_{c}^{-2}(\mathbf{Y}\circ\mathbf{\Pi}_{i:})\mathbf{C}_{i:}^T+\sigma_{r}^{-2}\mathbf{U}\mathbf{R}_{i:}\right]
	\end{gathered}
	\end{equation}

\textbf{Sampling} $\mathbf{R}$: {The sampling of $\mathbf{R}$ is given by $\mathcal{N}(\mathbf{R}_{ij}|\mu_{R_{ij}},1)$, where the mean $\mu_{R_{ij}}$ is:}

\begin{equation}
\small
\centering
\begin{gathered}
\label{eq:Rmean}
\mathbf{\mu}_{{R}_{ij}}=
\cfrac
{\mathbf{\Xi}_{ {ij}}(\mathbf{G}_{ {ij}}-0.5)\sigma_{R}^2+
	\mathbf{X}_{:i}^T\mathbf{U}_{:j}}
{\mathbf{\Xi}_{ {ij}}\mathbf{\Lambda}_{ {ij}}\sigma_{R}^2+1}
\end{gathered}
\end{equation}

\subsection{Predicting Missing Data}
By learning the latent matrices $\mathbf{{X,Y,U,}}$ and $\mathbf{{\xi}}$, $\mathbf{\hat{G}}=\sigma(X^TU)$ can be estimated. For the missing entry of 
$\mathbf{\acute{G}}_{ij}$, $\mathbf{\hat{G}}_{ij}$ is given by:
\begin{equation}
\small
\label{eq:gesimate}
{ \mathbf{\hat{G}}_{ij}}=\left\{\begin{matrix}
0 & if ((\xi_{ij}=0) \vee (\sigma{ (X_{:i}^{T}U_{:j}}) \preceq \delta_{G}))\\ 
1& else
\end{matrix}\right.
\end{equation}
where $\delta_{G}$ is a threshold for distinguishing links and non-links.
To approximate the missing values of $\mathbf{\acute{C}}_{ij}$, first we define the infection probability of the node $i$ in the cascade $j$ with $\mathbf{P}_{ij}$, and if $\mathbf{P}_{ij}  \succ  \delta_C$ then $\hat{\mathbf{C}}_{ij}$ is estimated by { $X_{:i}^{T}Y_{:j}$} with mapping the negative and the out of interval $[0,T]$ values to uninfected state:
\begin{equation}
\small
\begin{gathered}
\label{eq:cesimate}
{ 
z=X_{:i}^{T}Y_{:j}},\\
{ \mathbf{\hat{C}}_{ij}}=\left\{\begin{matrix}
0 & if ((\mathbf{P}_{ij} \preceq \delta_{C}) \vee (z \prec 0  \lor z \succ T))\\ 
z& else
\end{matrix}\right.
\end{gathered}
\end{equation}
The infection probability matrix is defined as $\mathbf{P}=\mathbf{\Pi}^T\mathbf{A}$, where $\mathbf{\Pi}$ is the mask diffusion matrix that we have already defined. $\mathbf{A}$ is a $N \times N$ infection transfer matrix, that each of its $(i,j)$ elements indicates the probability of infection propagation from node $i$ to node $j$ as Equation \eqref{eq:difftrans}, { where $|\{*\}|$ is the cardinality of set $*$.} {In other words, it represents the ratio of the number of cascades that node $i$ has been involved in before node $j$, in comparison to the total number of cascades each of them participated in.}
{ 
\begin{equation}
\small
\label{eq:difftrans}
\mathbf{A}_{i \rightarrow j}
=\cfrac
{| \{ (\mathbf{\acute{C}}_{i:}  \cap \mathbf{\acute{C}}_{j:}) 
	| (\mathbf{\acute{C}}_{i:} \prec \mathbf{\acute{C}}_{j:})
\} | }
{| \{ (\mathbf{\acute{C}}_{i:}  \cup \mathbf{\acute{C}}_{j:}) \} |}
\end{equation}}
From the diffusion observations, we obtain the probability that the simultaneous concurrence of two nodes $i$ and $j$ in cascades is due to the infection transmission from $i$ to $j$.\\
{ The DiffStru algorithm is presented as Algorithm \eqref{alg:overallalgorithm}.}

\begin{algorithm}
	\caption{DiffStru Algorithm}
	\label{alg:overallalgorithm}
	\SetAlgoLined
	\SetKwInOut{Input}{Input}
	\SetKwInOut{Output}{Output}
	\Input{Observed Matrix Cascade  $\mathbf{\acute{C}}$, Observed Matrix Graph  $\mathbf{\acute{G}}$,$\alpha_1,\alpha_2,\sigma_{C}^2,\sigma_{R}^2$
		, max iteration $T$
		, burn in $b$, thinning $k$, and latent factor $D$}
	\Output {Inferred Matrix Cascade $\mathbf{\hat{C}}$, Inferred Matrix Graph $\mathbf{\hat{G}}$}
	$(N,N)=$shape ($\mathbf{G}$)\;
	$(N,M)=$shape ($\mathbf{C}$)\;
	Initialize $\mathbf{{W}_Y}^{-1},\mathbf{{W}_X}^{-1},\mathbf{{W}_U}^{-1},$ with Equation \eqref{eq:Xcovaraince} or identity matrix \\
	Initialize $\mathbf{{X^{D \times N}},R^{N \times N},Y^{D \times M},U^{D \times N},\Lambda^{N \times N},\mu^{N \times N}}$ randomly
	Initialize binary mask $\mathbf{\Xi}^{N \times N}$ and $\mathbf{\Pi} ^ {N \times M}$ from $\mathbf{\acute{G}}$ and $\mathbf{\acute{C}}$\;
	\For{t=1,...,T}{
		Sample $\mathbf{R}^{t}$ from normal with \eqref{eq:Rmean}\\
		Sample each element of ${\mathbf{\Xi}_{ij}}^{t}$ from \eqref{eq:FinalSamplingE}\\
		Sample $\mathbf{X}^{t}$ from \eqref{eq:Xcovariance} or \eqref{eq:Xcovariance2} \\
		Sample $\mathbf{U}^{t}$ from \eqref{eq:Ucovariance} or \eqref{eq:Ucovariance2}\\
		Sample $\mathbf{Y}^{t}$ from \eqref{eq:Ycovariance} or \eqref{eq:Ycovariance2}\\
		Sample each element of { ${\mathbf{\Lambda}_{ij}^{t}}$} from \eqref{eq:Samplinglambda}\\
		Sample $\mathbf{\mu}^{t}$ from \eqref{eq:Samplingmu}\\
	}
	$\mathbf{X}=\cfrac{k}{T-b}\displaystyle\sum\limits_{i=1}^{\frac{T-b}{k}}{\mathbf{X}^{b+ki}}$,
	$\mathbf{Y}=\cfrac{k}{T-b}\displaystyle\sum\limits_{i=1}^{\frac{T-b}{k}}{\mathbf{Y}^{b+ki}}$\\
	$\mathbf{U}=\cfrac{k}{T-b}\displaystyle\sum\limits_{i=1}^{\frac{T-b}{k}}{\mathbf{U}^{b+ki}}$,
	$\mathbf{\Xi}=\cfrac{k}{T-b}\displaystyle\sum\limits_{i=1}^{\frac{T-b}{k}}{\mathbf{\Xi}^{b+ki}}$
	
	Compute $\mathbf{\hat{G}}$ from \eqref{eq:gesimate}\\
	Compute $\mathbf{\hat{C}}$ from \eqref{eq:cesimate}


\end{algorithm}

\subsection{{Computational Complexity}}
{The most time-consuming part of the algorithm is sampling three latent factors. In a parallel environment, the time complexity of sampling of { variable $\mathbf{U}$ as Equation} \eqref{eq:Ucovariance2} is $O(D^3+D^2N)$, { variable $\mathbf{X}$ based on Equation} \eqref{eq:Xcovariance2} is $O(D^3+D^2N+D^2M)$, and { variable $\mathbf{Y}$ from Equation} \eqref{eq:Ycovariance2} is equal to $O(D^3+D^2N)$. Hence, in $T$ iterations, the total complexity of the algorithm is equal to $O(TD^2 (D+N+M))$, which is linear with respect to $N$, $M$ and $D \prec\prec M,N$.}

\section{Numerical and Simulation Results}\label{sec:numerical and simulation results}
For supporting different network scales, we implement DiffStru in a distributed framework named Ray \cite{moritz2018ray}, which is an open-source framework for scaling up Python applications from single machines to large clusters. Depending on the amount of input data and resources required in each step of the algorithm's iteration, new machines can enter the cluster during the run of the method.
{ We also used Ray for tuning hyper-parameters of our method and the rest of the works with which we have compared}.
{ We ran all experiments on a server with an Intel\textsuperscript{®} Core™ i9-7980XE Extreme Edition Processor equipped with 128GB of RAM and two NVIDIA GeForce GTX 1080 TI, running Ubuntu 18.04.}
Since no related work addresses joint completion of the missing parts of diffusion and structure, we separately compare DiffStru against different {  link prediction, network inference, and cascade
correction and prediction methods.}

\textbf{Structure part competitors}: We chose Adamic Adar (\textbf{AA}) \cite{adamic2003friends}, Resource Allocation (\textbf{RA}) \cite{ou2007power}, and Common Neighbor (\textbf{CN}) \cite{lorrain1971structural} as classical link prediction methods as well as the recent fusion matrix factorization method (\textbf{FPMF}) \cite{MFLink2018}. The classical approaches calculate the similarity weight of any unlinked pairs of nodes. Unobserved links can be obtained by sorting the weights in ascending order and cutting the top pairs. FPMF is a matrix factorization method that fuses some asymmetric and symmetric topological metrics with an adjacent structure matrix to learn the latent factor of nodes with gradient methods. {  Link-Corrected Prediction Algorithm\textbf{LCPA} \cite{chai2022network} is a maximum likelihood method for estimating hidden links. These works are based on the incomplete structure of networks as inputs.}\\
{ 
From the network inference category, we selected the \textbf{Netrate} \cite{gomez2013modeling}, \textbf{DANI} \cite{ramezani2017dani}, and \textbf{REFINE} \cite{kefato2019refine} methods that find hidden links using fully observed diffusion information. We used partially observed cascades as the input of these methods}.

\textbf{Cascade part competitors}: There is no work for mining the omitted infected nodes in the diffusion category by estimating their infection time. To compare this part of DiffSru, we chose a cascade prediction technique called \textbf{DeepDiffuse} \cite{DeepDiffuse2018}. DeepDiffuse tries to predict the next node in a cascade by ranking the possible candidates by estimating a single time. However, this method does not pay attention to missing data and aims to predict the next step of the cascade sequence by assuming that the observed steps are complete. Therefore, to compare DeepDiffuse with DiffSru, we should input the cascade sequence up to any missing step point to the model to obtain the prediction. Then we can compare the error of the predicted next infection time. As a baseline, we fit a polynomial regression model (with degrees one and two, named Reg-1 and Reg-2) on a sorted cascade sequence. Then for any missing node $k$, which is infected after node $i$ and before node $j$, we find the related time from the learned regression model using the mean indexes of node $i$ and $j$ in the sequence. We provide more prior knowledge for both DeepDiffuse and baseline methods compared to DiffStru. Both methods know the position of two nodes in a cascade where the missing has happened. Therefore, computing the missed infection time is approximated relative to the time of the previous node. However, DiffStru does not utilize this information. {  While DeepDiffuse predicts the next infected node with its infection time, other cascade prediction methods included in Section \eqref{sec:related work} do not predict time. Since we aim to detect missing nodes with their infection time, DeepDiffuse is the only comparison candidate from the "cascade correction and prediction" category.} For more experiments, we used \textbf{NetFill} \cite{sundareisan2015hidden}, which finds the missing nodes in a cascade without their infection time by using the entire structure of the network. The only way to compare DiffStru and NetFill is based on the inferred number of missed infection nodes.

{ Joint Weighted Nonnegative Matrix Factorization (\textbf{JWNMF}) \cite{tang2023joint} is another link prediction method using non-negative matrix factorization and binary attributes of nodes. By defining the cascade participle as a node attribute, we construct this attribute matrix from partially observed cascades. Whenever a node is infected in a cascade, its related attribute is $1$, otherwise $0$. By doing this, we make \textbf{JWNMF} the closest related work with the same inputs as DiffStru.}

{ 
Our experiments were conducted using the official code provided by the authors of all related works, except REFINE and JWNMF (REFINE and JWNMF codes were not available). We implemented their methods ourselves based on the explanations provided in their papers. We used ray hyper-parameter tuning for each of these two methods to achieve the best results. Our implementations of REFINE\footnote{https://github.com/AryanAhadinia/REFINE} and JWNMF\footnote{https://github.com/AryanAhadinia/JWNMF} are available on GitHub. }

%
%
%
%
%
\subsection{Description of the Datasets}\label{sec:dataset}
\textbf{Synthetic}:
We generated synthetic directed graphs resembling real social networks independent of the proposed model. The LFR (Lancichinetti-Fortunato-Radicchi) network is a benchmark having interesting real-world features such as community structure and power-law degree distribution \cite{LFRbenchmark2008}. We generated three different artificial LFR networks having $50$, $100$, and $400$ nodes with parameters: mixing parameter 0.1, degree sequence exponent 2, community size distribution exponent 1, and zero number of overlapping nodes. Then, we simulated different independent cascades on each graph using the method described in \cite{gomez2012inferring}. The transmission model was exponentially distributed with parameters: alpha=1, a mixture of exponential=1, and beta=0.5.\\ 
\textbf{Real world}: As stated earlier, data loss in real-world datasets is unavoidable. Thus, creating a test scenario on these datasets will cause an extreme lack of information, and no proper ground truth exists for evaluation. Therefore, we consider a dense part of these networks (by choosing a big community) with a density of about $0.1$ to test and perform data deletion scenarios. Therefore, 
{we applied this approach to two real datasets:}
A Twitter dataset during October $2010$ with a network of $6642$ links between followers 
and time sequence of retweeting between $2144$ activity of users \cite{hodas2014simple}. Also, a Memestracker dataset with $4828$ memes propagation between websites during April 2009 with a $41404$ graph connection
\cite{leskovec2009meme}.

\subsection{Evaluation Metrics}\label{sec:metrics}
To evaluate the results of DiffStru, we compared its performance from structural and diffusion points of view. The observed $\mathbf{\acute{G}}$ and $\mathbf{\acute{C}}$ are the training sets, and we treat them as known data, while the missing parts of $\mathbf{{G}} \backslash \mathbf{\acute{G}}$ and $\mathbf{{C}} \backslash \mathbf{\acute{C}}$ are testing sets, and no knowledge about these sets is used in learning our model. If the output of models is represented with $\mathbf{\hat{G}}$ and $\mathbf{\hat{C}}$, then we evaluate the reconstruction of estimated matrices and accuracy of different aspects with the following metrics:\\
\textbf{SRE}: With Signal Reconstruction Error (SRE) \cite{iordache2011sparse}, we can evaluate the success rate in reconstructing the ground-truth matrix - the higher this metric, the lower the recovery noise ratio to the original matrix. For the ground-truth matrix $\mathbf{Z}$, if the estimated matrix is $\mathbf{\hat{Z}}$, SRE metric is approximated as \cite{iordache2011sparse}:
\\
\begin{equation}
\small
\label{eq:sre}
SRE=\frac{\left\lVert \mathbf{Z}\right\lVert^2_2}
{\left\lVert \mathbf{\hat{Z}}-\mathbf{Z}\right\lVert^2_2}
\end{equation}
\textbf{AUC}: Since we face an unbiased binary classification in reconstructing the structural information (matrix $\mathbf{G}$), the area under the receiver operating characteristic curve is a useful metric for evaluation, which is independent of the threshold setting. AUC measures the probability of randomly selecting linked and unlinked pairs of nodes and checking if the probability of linked pairs is higher than unlinked pairs. The higher AUC value indicates the better performance of the model in classifying the two classes.\\
\begin{equation}
\small
\begin{gathered}
\label{eq:AUC}
\tiny
AUC= 
\frac{\sum_{(a,b),(c,d) \in ({\mathbf{{G}} \backslash \mathbf{\acute{G}}})}
	\left[ \mathbb{I}( \mathbf{\hat{G}}(a,b) \succ \mathbf{\hat{G}}(c,d))+0.5( \mathbf{\hat{G}}(a,b)= \mathbf{\hat{G}}(c,d)) \right]
	\times \mathbb{I}(\mathbf{G} (a,b)=1, \mathbf{G}(c,d)=0)}
{\sum_{(a,b),(c,d) \in ({\mathbf{{G}} \backslash \mathbf{\acute{G}}})}{\mathbb{I}(\mathbf{G} (a,b)=1, \mathbf{G}(c,d)=0)}}
\end{gathered}
\end{equation}
\textbf{Precision, Recall, and F-measure}: By applying a threshold on probabilities of matrix $\mathbf{\hat{G}}$, we can map the elements to zero and one values. The precision and recall are defined below, and F-measure is the weighted harmonic mean of recall and precision.
\begin{equation}
\small
\begin{gathered}
\label{eq:precision}
Precision=\frac{\mathbf{\acute{G}}=0\wedge\mathbf{G}=1 \wedge \mathbf{\hat{G}}=1}
{\mathbf{\hat{G}}=1 \wedge \mathbf{\acute{G}}=0}, Recall=\frac{\mathbf{\acute{G}}=0 \wedge \mathbf{G}=1 \wedge \mathbf{\hat{G}}=1}
{\mathbf{G}=1 \wedge \mathbf{\acute{G}}=0}
\end{gathered}
\end{equation}
\textbf{Accuracy}: Is a metric for measuring the portion of correctly classified pairs:
\begin{equation}
\small
\begin{gathered}
\label{eq:acc}
ACC=\cfrac{(\mathbf{\acute{G}}=0 \wedge \mathbf{G}=1 \wedge \mathbf{\hat{G}}=1)
	+
	(\mathbf{G}=0 \wedge \mathbf{\hat{G}}=0)
}
{\mathbf{\acute{G}}=0}
\end{gathered}
\end{equation}\\
\textbf{MCC}: The Matthews Correlation Coefficient metric \cite{matthews1975comparison} is suitable for measuring the quality of binary ill-balanced classification, which utilizes the ratios of true positives, true negatives, false positives, and false negatives. When the value of this metric is near $1$, the estimated value is closer to reality, while the value closer to $0$ shows the estimator prediction is close to random, and the $-1$ value indicates a complete disagreement.\\
\textbf{MAP$@K$}: The Mean Average Precision is utilized for measuring the performance of DeepDiffuse \cite{DeepDiffuse2018}. If we have $\#A$ number of missing activities, the ground truth of these test samples has one value, while DeepDiffuse outputs an ordered list of nodes. MAP$@K$ means averaging overall prediction of $\#A$ samples based on the $@K$ top first of the algorithm output list $\frac{\sum\limits_{i = 1}^{\#A}{[\mathbb{I}(r_i \preceq K)(\frac{1}{r_i})}]}{\#A}$\\
\textbf{RMSE}: We measure the Root Mean Square Error (RSME) between $\mathbf{C}$ and $\mathbf{\hat{C}}$ for the timestamp of observed training data and test unobserved nodes.
There is no fixed threshold limit for RMSE, and the lower value indicates a better fit.
\subsection{Experiments Setting}\label{sec:setting}
The first step toward setting inputs of experiments is simulating the missing information in all datasets. Missing data can occur at random events, which leads to an unbiased analysis. However, in some real cases, data missing may happen because of sampling techniques or some limitations of APIs that are not random. For example, we cannot gather all the connections of a high-degree node that directs us to the missing data. Since we do not have any assumption about deletion in the proposed method, we ran different experiments for two types of missing data.\\
\textbf{Random missing}: We eliminate data with uniform distributions to simulate random missingness. For this purpose, we first choose a remove rate $\theta \in [0,1]$. Then for each link of structure or activity in the diffusion, a random value $\tau$ is generated from a uniform distribution on $[0,1]$, and if ($\tau \prec \theta$), we remove that data.\\
\textbf{Non-random missing}: For a non-random scenario, we randomly remove the links between nodes whose outdegree is higher than five. Besides, we calculate the activities of each node, and activity removal is done randomly over the set of nodes with more than five participants. 
By default, runs were over $1000$ iterations with burn-in of $900$ and thinning of $1$. If there is a change in the setting of any scenario, we have listed the new settings.

\subsection{Hyper-parameters Setting}\label{sec:hypersetting}
The link prediction between any two nodes is correlated with the structural properties of nodes so that the more similar structural relations of two nodes will lead to more likelihood of a link between them. Furthermore, the cascades that have infected more similar nodes are more alike in their diffusion path. Suppose that $ \mathbf{\Theta}^{X}_{ij}$ is a measure of the structural similarity between two nodes $i$ and $j$ in a network, then the larger value of $ \mathbf{\Theta}^{X}_{ij}$ indicates the lower distance between $\mathbf{X}_{:i}$ and $\mathbf{X}_{:j}$. Therefore, for any pair of nodes, we look for the minimal value of the following equation:
\begin{equation}
\small
\begin{gathered}
\label{eq:Xcovaraince}
\sum\limits_{i = 1}^N {\sum\limits_{j = 1}^N {\mathbf{\Theta}^{X}_{ij}\left\| {\mathbf{X}_{:i} - \mathbf{X}_{:j}} \right\|} _2^2}
=
\sum\limits_{i = 1}^N {\sum\limits_{j = 1}^N \sum\limits_{d=1}^D ({\mathbf{\Theta}^{X}_{ij}{\mathbf{X}^2_{di}} 
		-
		\mathbf{\Theta}^{X}_{ij}{\mathbf{X}_{di}}{\mathbf{X}_{dj}}})}	
=\sum\limits_{d=1}^D{{\mathbf{X}}^T_{d:}}{(\mathbf{\zeta}^{X}- \mathbf{\Theta}^{X})}{\mathbf{X}_{d:}}
\end{gathered}
\end{equation}
where $\mathcal{L}^{X}={(\mathbf{\zeta}^{X}- \mathbf{\Theta}^{X})}$ is a Laplacian matrix with
$\mathbf{\zeta}^{X}_{ii}=\sum\limits_{j=1}^{N}\mathbf{\Theta}^{X}_{ji}$
is a diagonal matrix. This property of $\mathbf{X}$ is equivalent to multivariate Gaussian distribution with zero mean and inverse covariance matrix ${\mathbf{W}_X}^{-1}=\mathcal{L}^{X}$. Similarly, we can model $\mathbf{U}$ and $\mathbf{Y}$. 
To initialize the covariance matrix of hyper-parameters, we can set it to an identity matrix without modeling the correlation of nodes and cascades (independent priors) or considering the relation of these components (correlated priors). For dependent priors, we use the similarity of nodes as the number of common neighbors in the partially observed structure network for $\mathbf{\Theta}^{X}$ and $\mathbf{\Theta}^{U}$. In contrast, the similarity of cascades is defined by the ${ij}$ entry of $\mathbf{\Theta}^{Y}$, which is the number of common nodes in $i$-th and $j$-th cascades.
Other hyper-parameters are set as, in Equation \eqref{eq:cesimate} where $\delta_G=0.5$ and $\delta_C$ in Equation \eqref{eq:gesimate} is set as the mean value of all $\mathbf{P}$ elements. Moreover, ${\sigma_C}^2=1$, ${\sigma_R}^2=1$, $D=8$, $\alpha_{1}=0.2$, and $\alpha_{2}=0.3$. {  All the elements of matrix $\mathbf{X}$, $\mathbf{U}$, and $\mathbf{Y}$ have been initialized by sampling from the Rectified Normal Distribution so that all elements are positive.}
\subsection{Model Analysis}\label{sec:modelanalysis}
\begin{figure}[ht]
    \centering
    \includegraphics[width=1\textwidth]{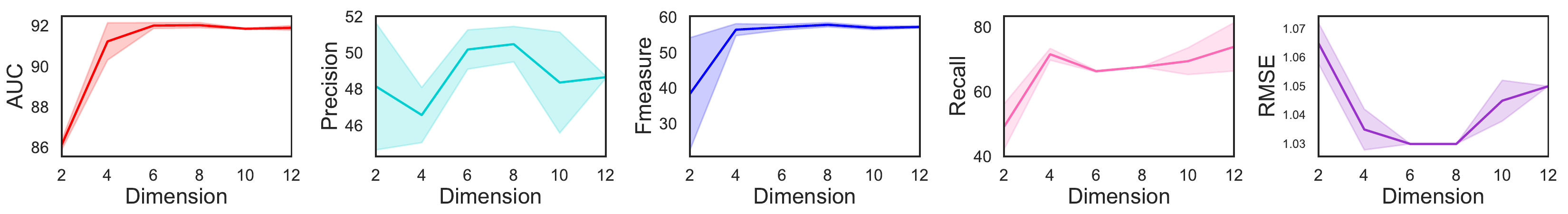}
    \caption{{  Comparing structural and diffusion metrics against different values of latent space in the LFR100 dataset.}}
	\label{fig:latentfactor}
\end{figure}

In this section, we verify the impact of different parameters on the performance of DiffStru and focus on its ability to solve problems in various situations. Some experiments on synthetic LFR networks are intended to answer the following questions:
%
%
%
\\(1) What is the best size of the latent space (hyper-parameter $D$)? \\
Learning the latent factors of the model in the smaller dimension will reduce the complexity of inference calculations. On the other hand, the model focuses on completing the structure and diffusion matrices simultaneously. While the measurement metrics for completing these matrices differ, we scan for a lower dimension where the best results can be achieved in both structure ad diffusion spaces. Because of random initialization, we ran the model multiple times and reported the mean and standard deviation of metrics, as shown in Fig. \eqref{fig:latentfactor}. Here, the different values of latent factors are analyzed, and $D=8$ is the best choice with a higher mean value and lower standard deviation in terms of AUC, precision, and F-measure for mining missing links, as well as lower mean and standard deviation for RMSE in finding missing activities. 

\begin{figure}[ht]
	\begin{center}
		\begin{subfigure}{0.45\textwidth}
			\centering\includegraphics[width=1\textwidth]{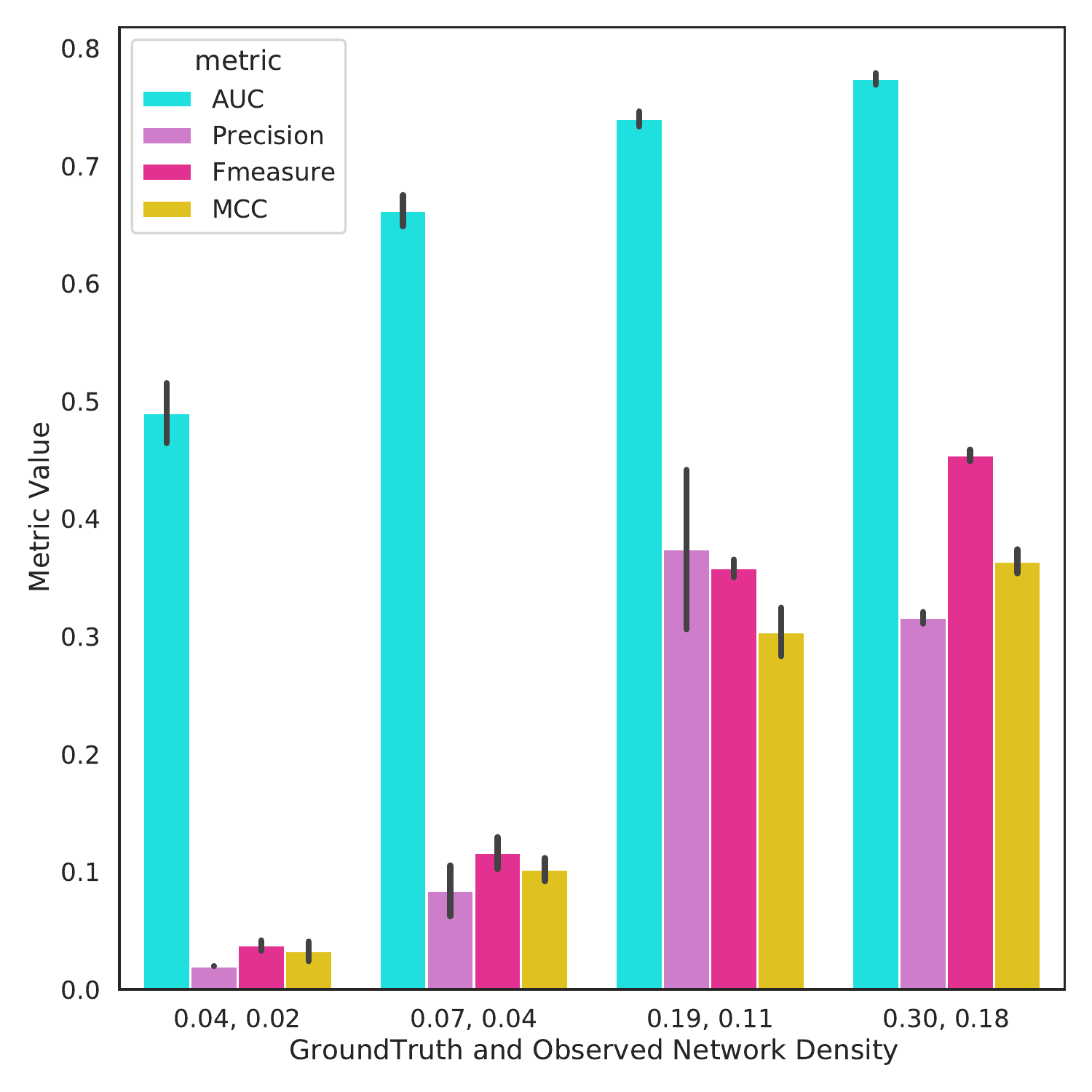}
			\caption{Structure}
			\label{fig:densitystru}
		\end{subfigure}
		\begin{subfigure}{0.45\textwidth}
			\centering\includegraphics[width=1\textwidth]{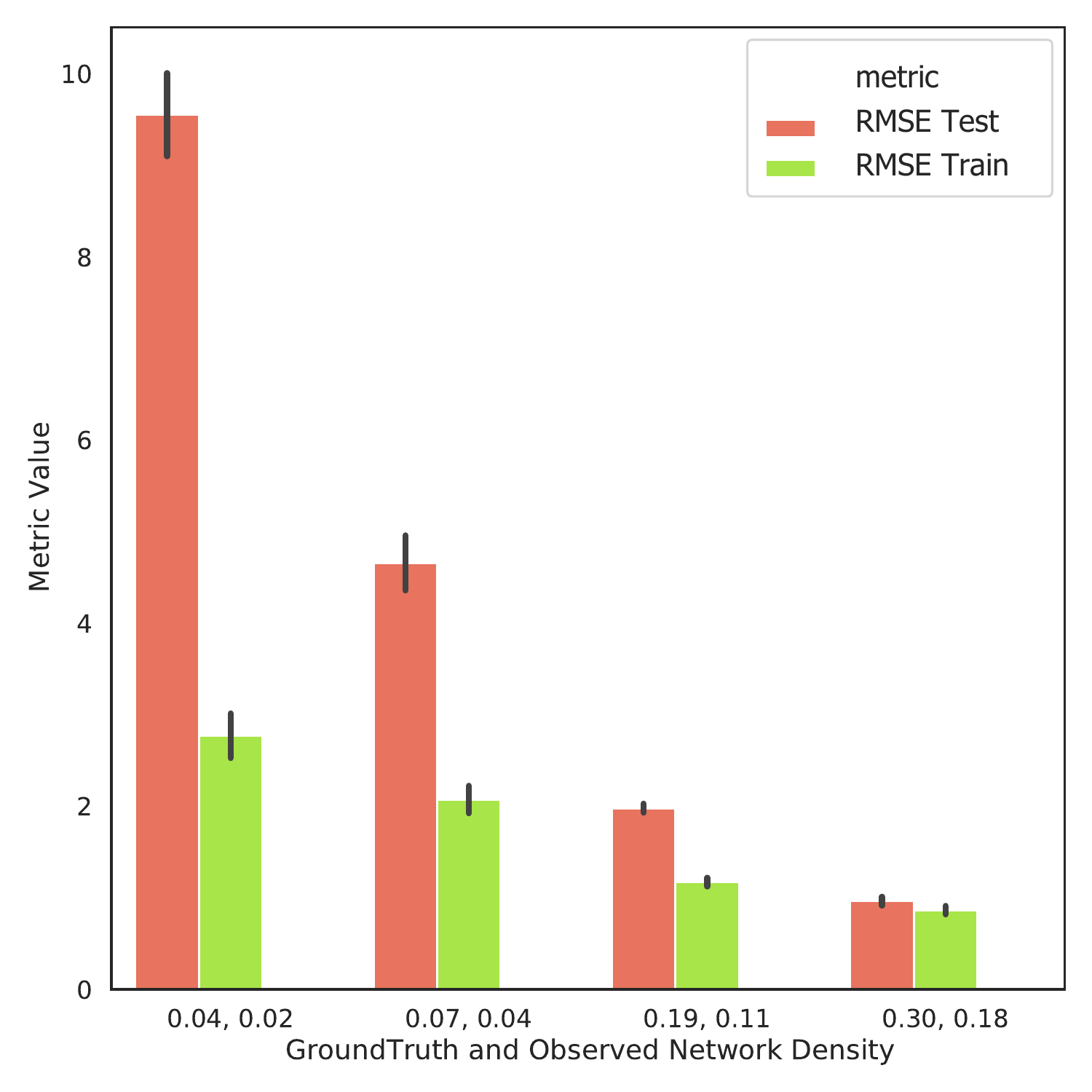}
			\caption{Diffusion}
			\label{fig:densitycas}
		\end{subfigure}
		\caption
		{{Impact of structure network density on LFR100 dataset. The horizontal axis of the plot represents the density of the ground truth and observed network. The input to the models is the observed network, which is less dense due to partially observed data (only $60\%$ of links are observed).}
		}
		\label{fig:acc}
	\end{center}
\end{figure}

\begin{figure}[ht]
	\begin{center}
		\begin{subfigure}{0.45\textwidth}
			\centering\includegraphics[width=1\textwidth]{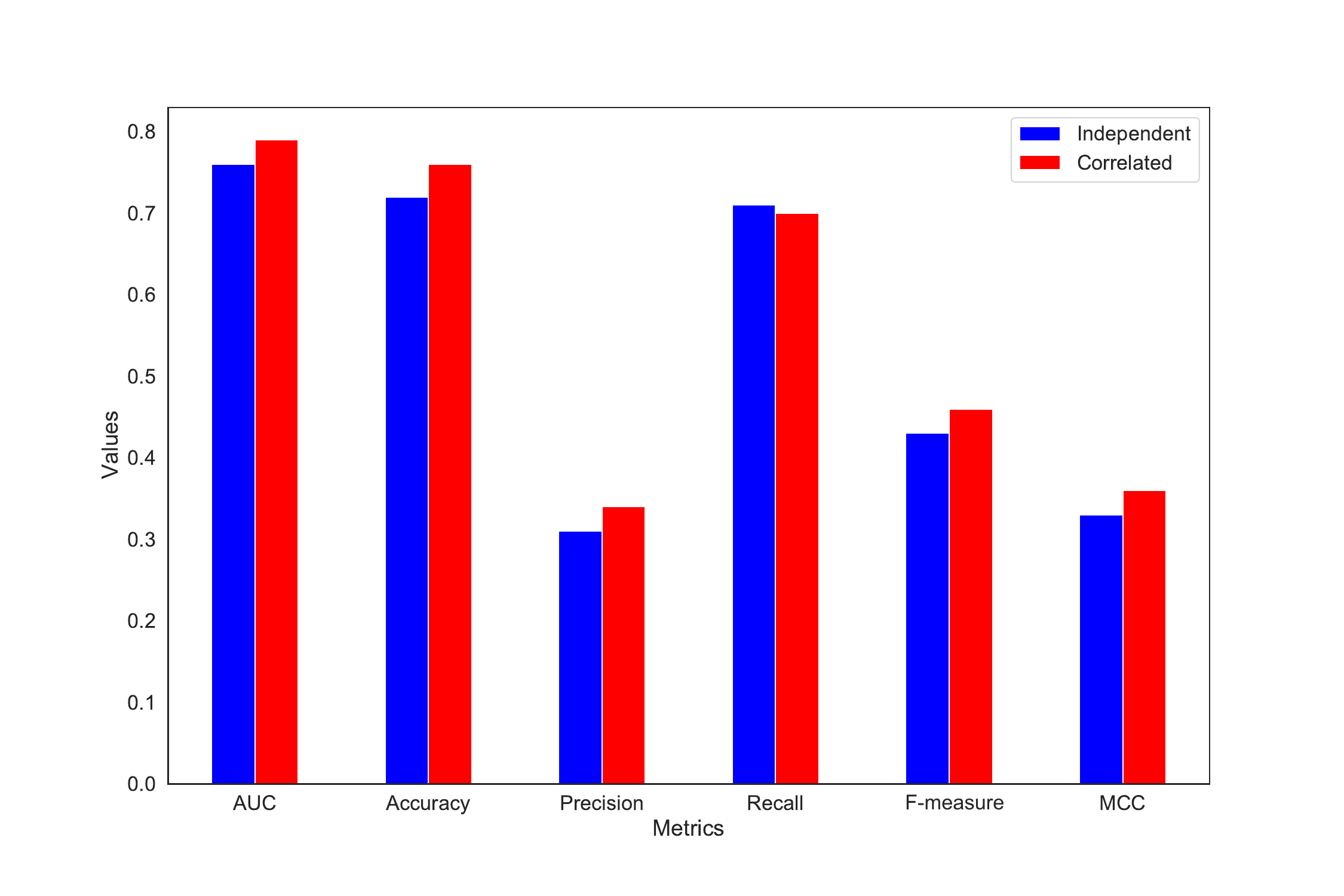}
			\caption{Structure}
			\label{fig:privateuser-stru}
		\end{subfigure}
		\begin{subfigure}{0.45\textwidth}
			\centering\includegraphics[width=1\textwidth]{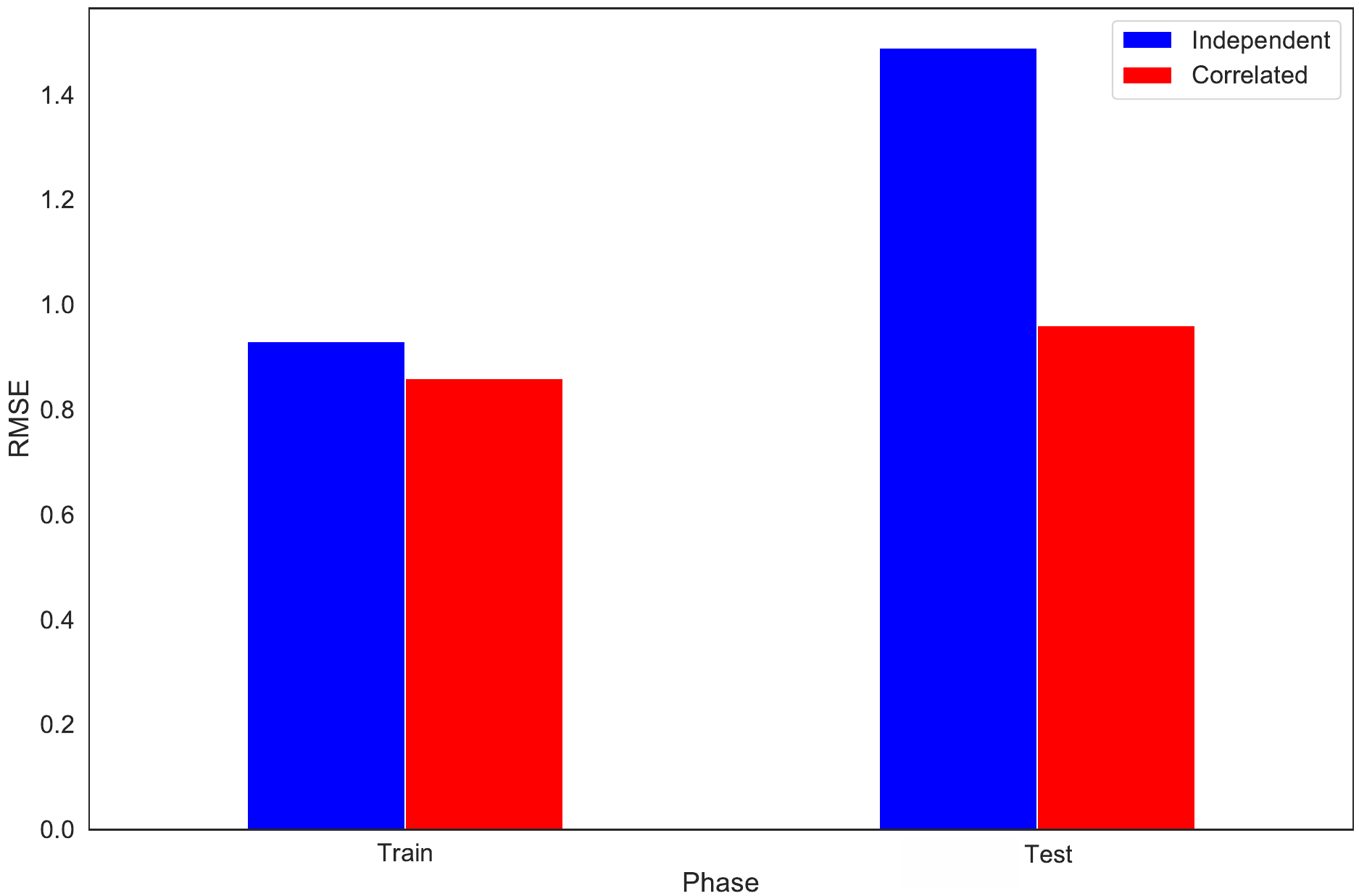}
			\caption{Diffusion}
			\label{fig:privateuser-cas}
		\end{subfigure}
		\caption{Effect of using correlated values for initializing the covariance matrix of prior distributions in LFR50 dataset.}
		\label{fig:privateuser}
	\end{center}
\end{figure}

\begin{figure}[ht]
	\begin{center}
		\begin{subfigure}{0.3\textwidth}
			\centering\includegraphics[width=1\textwidth]{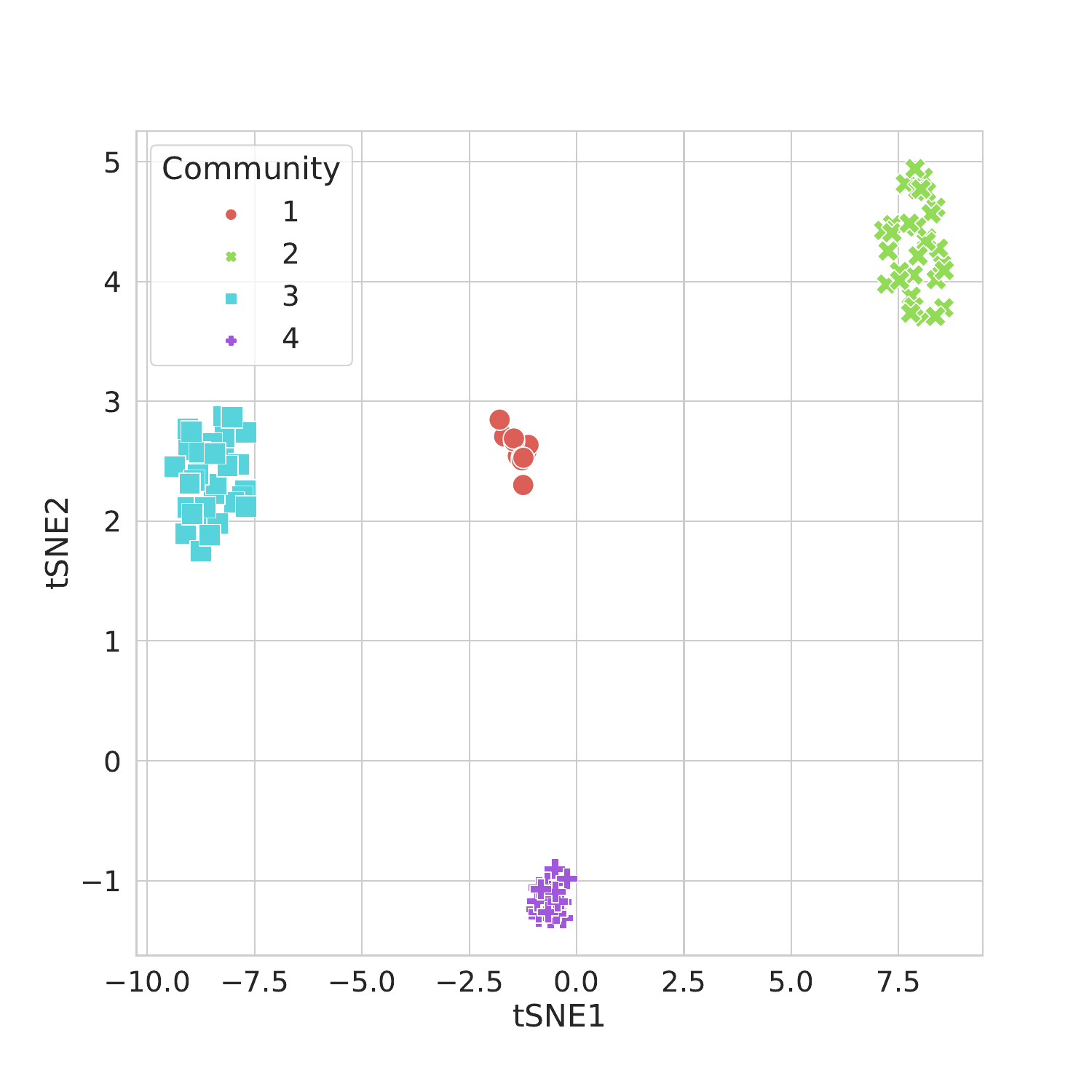} 
			\caption{\smaller {LFR100-$X$ matrix}}
			\label{fig:LFR100-X}
		\end{subfigure}
		\begin{subfigure}{0.3\textwidth}
			\centering\includegraphics[width=1\textwidth]{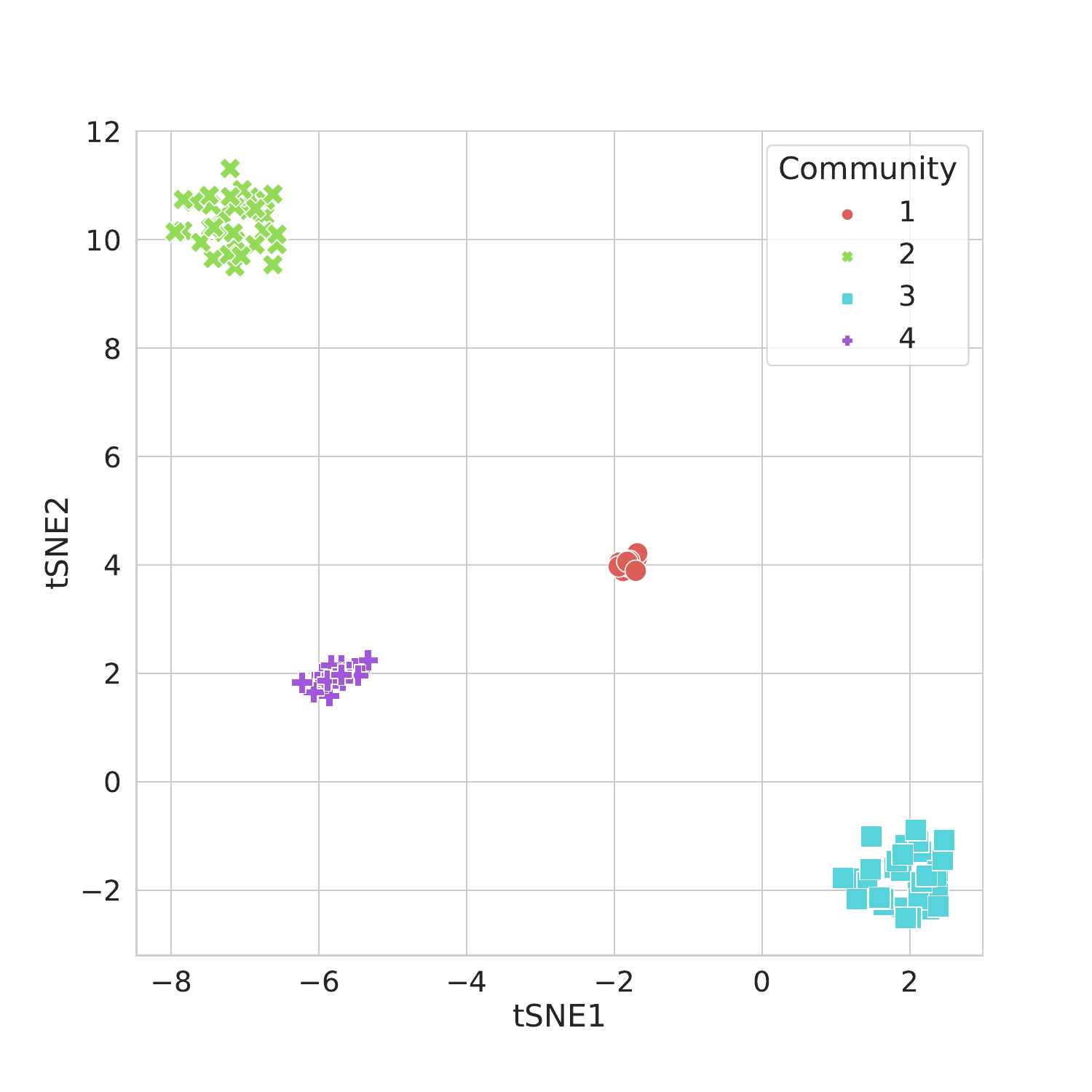} 
			\caption{\smaller {LFR100-$U$ matrix}}
			\label{fig:LFR100-U}
		\end{subfigure}
		\begin{subfigure}{0.3\textwidth}
			\centering\includegraphics[width=1\textwidth]{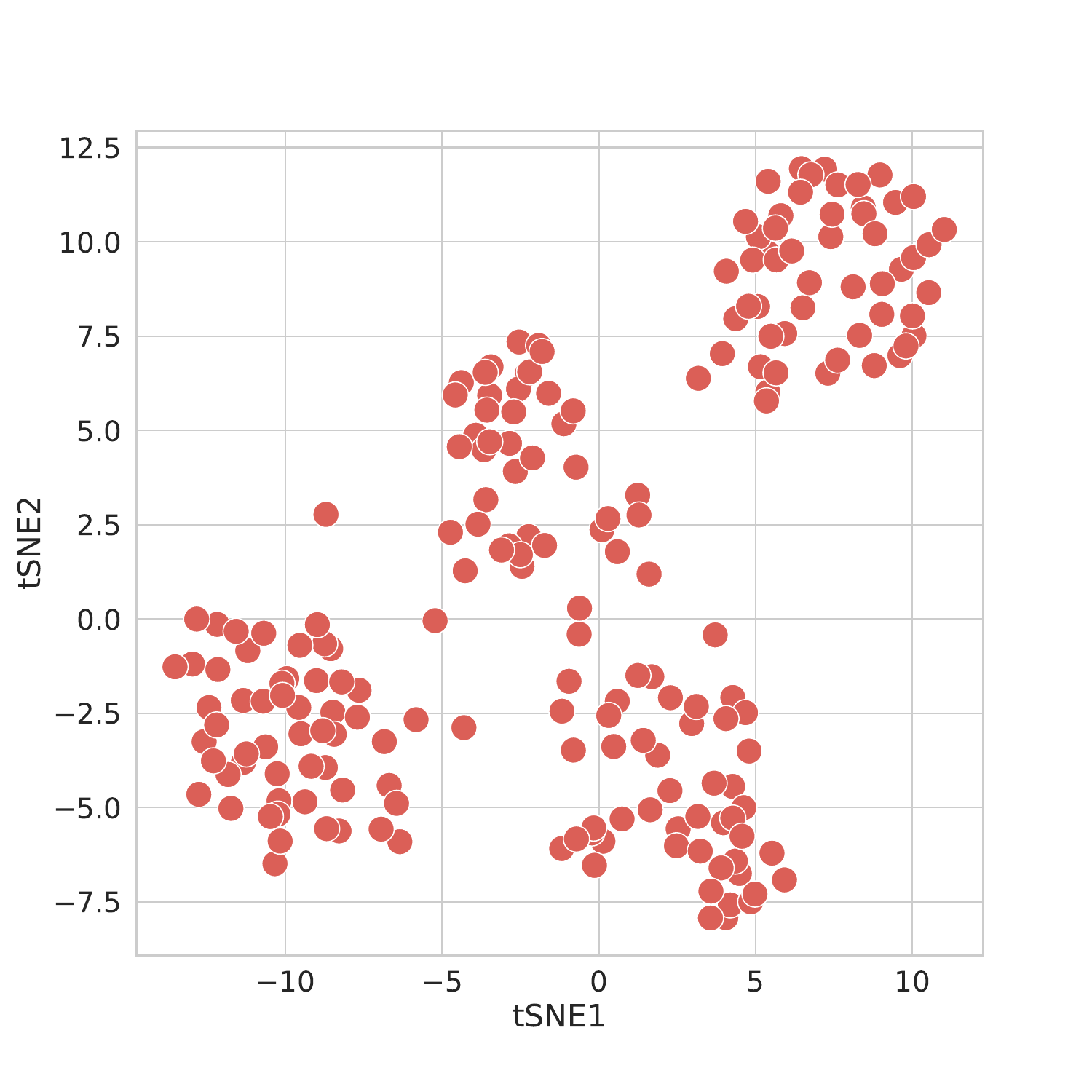} 
			\caption{\smaller {LFR100-$Y$ matrix}}
			\label{fig:LFR100-Y}
		\end{subfigure}
		\begin{subfigure}{0.3\textwidth}
			\centering\includegraphics[width=1\textwidth]{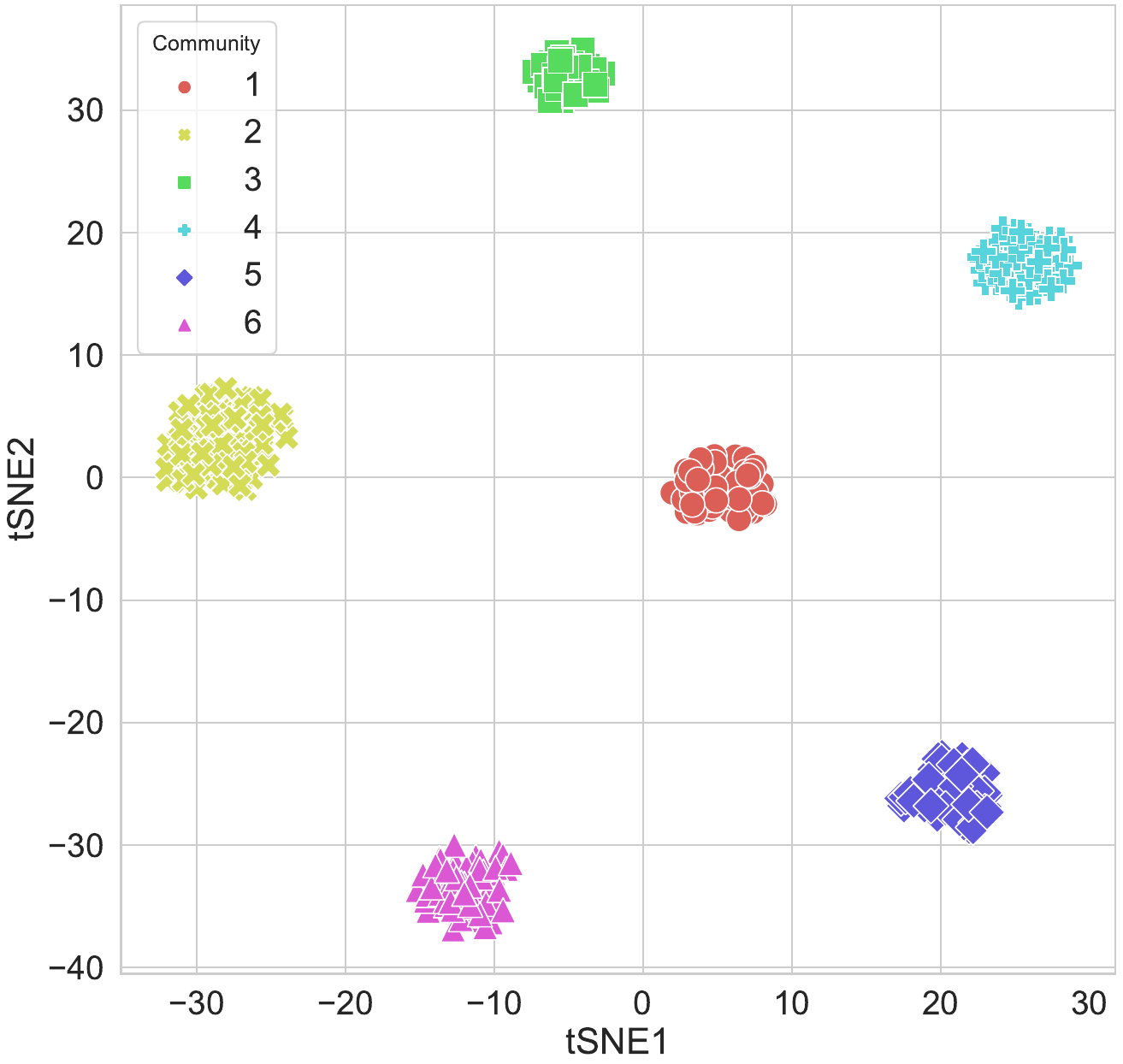}
			\caption{\smaller {LFR400-$X$ matrix}}
			\label{fig:LFR400-X}
		\end{subfigure}
		\begin{subfigure}{0.3\textwidth}
			\centering\includegraphics[width=1\textwidth]{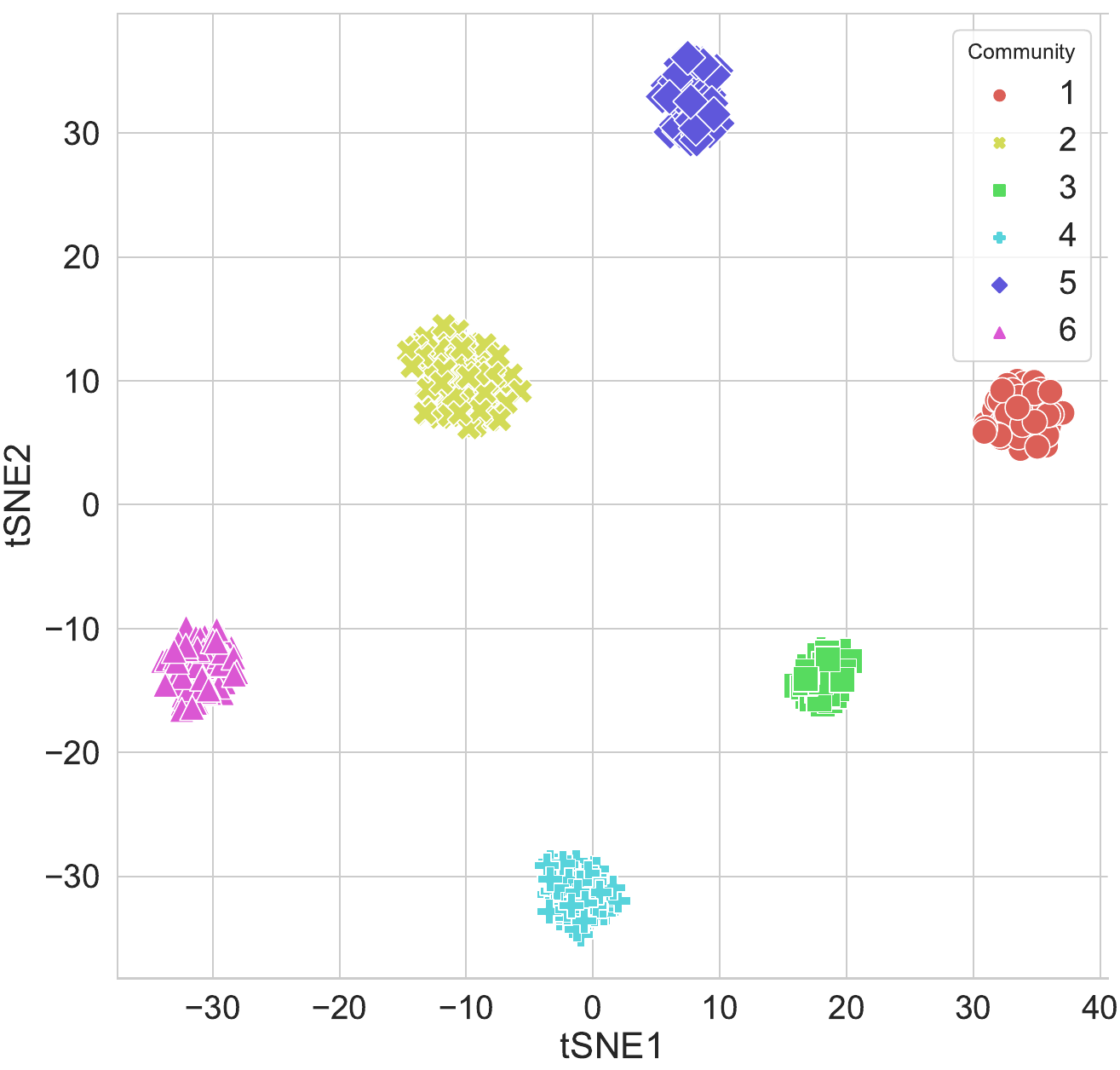}
			\caption{\smaller {LFR400-$U$ matrix}}
			\label{fig:LFR400-U}
		\end{subfigure}
		\begin{subfigure}{0.3\textwidth}
			\centering\includegraphics[width=1\textwidth]{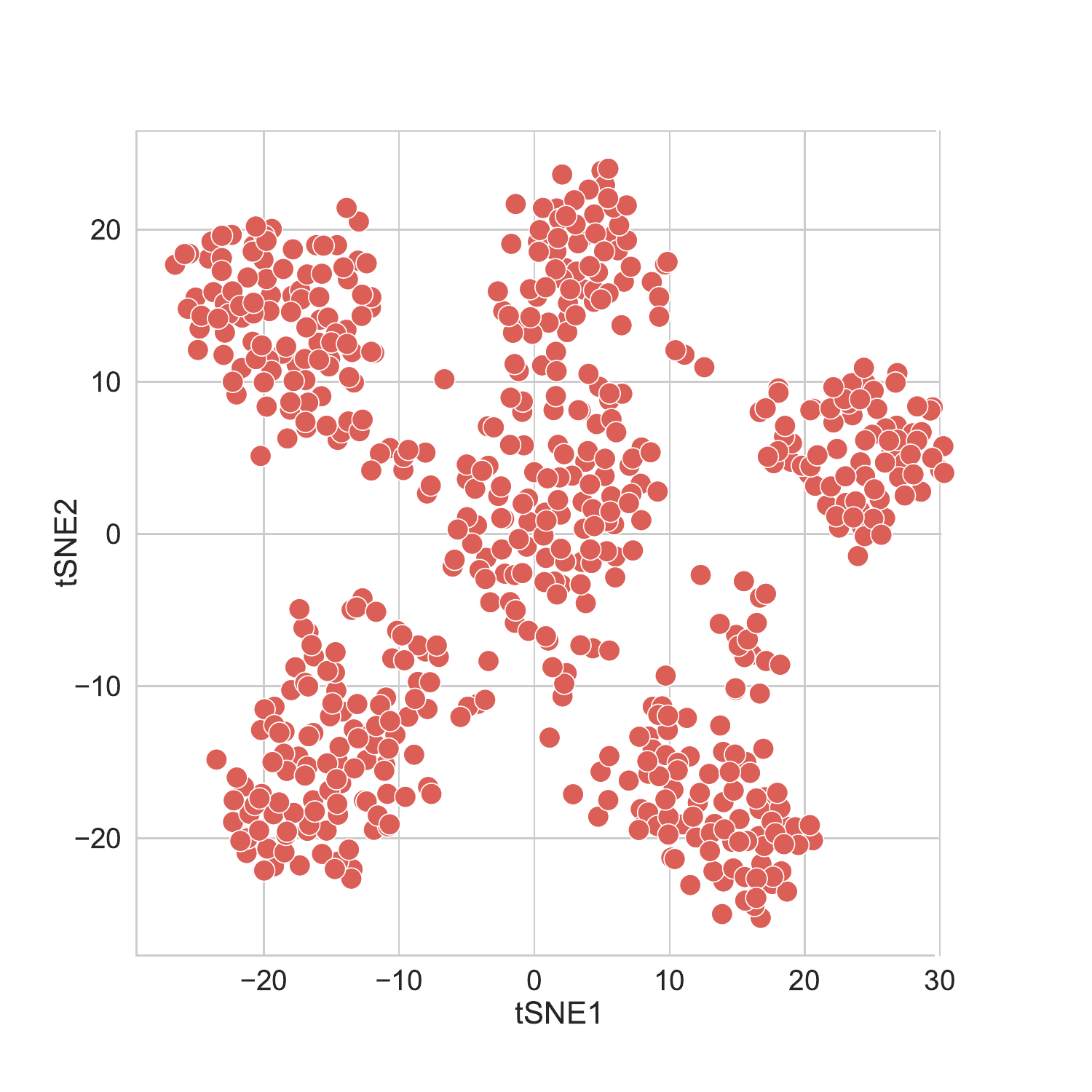}
			\caption{\smaller {LFR400-$Y$ matrix}}
			\label{fig:LFR400-Y}
		\end{subfigure}
		\caption{Visualization of learned latent feature matrices.}
		\label{fig:tsnecommunity}
	\end{center}
\end{figure}

\begin{figure}[ht]
    \centering
    \includegraphics[width=1\textwidth]{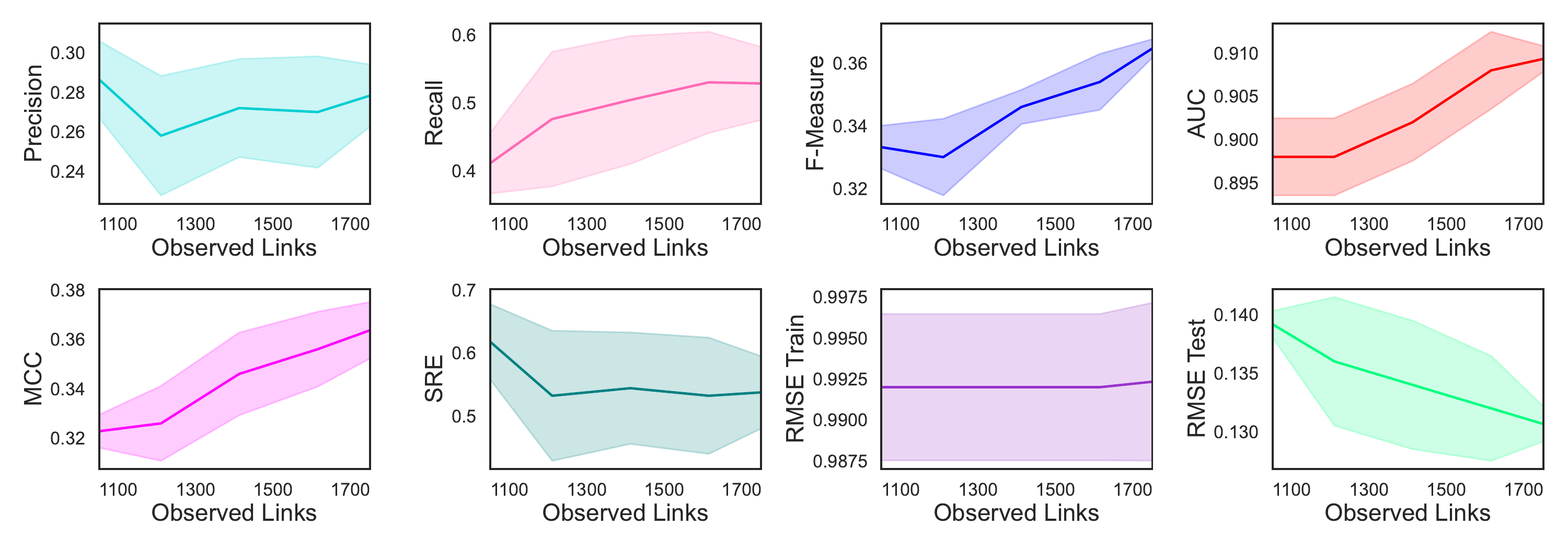}
    \caption{{  Impact of link removal rate on the link inferring and diffusion matrix results for LFR100 dataset.}}
    \label{fig:removelink}
\end{figure}

\begin{figure}[b]
	\begin{center}
		\begin{subfigure}{0.3\textwidth}
			\centering\includegraphics[width=1\textwidth]{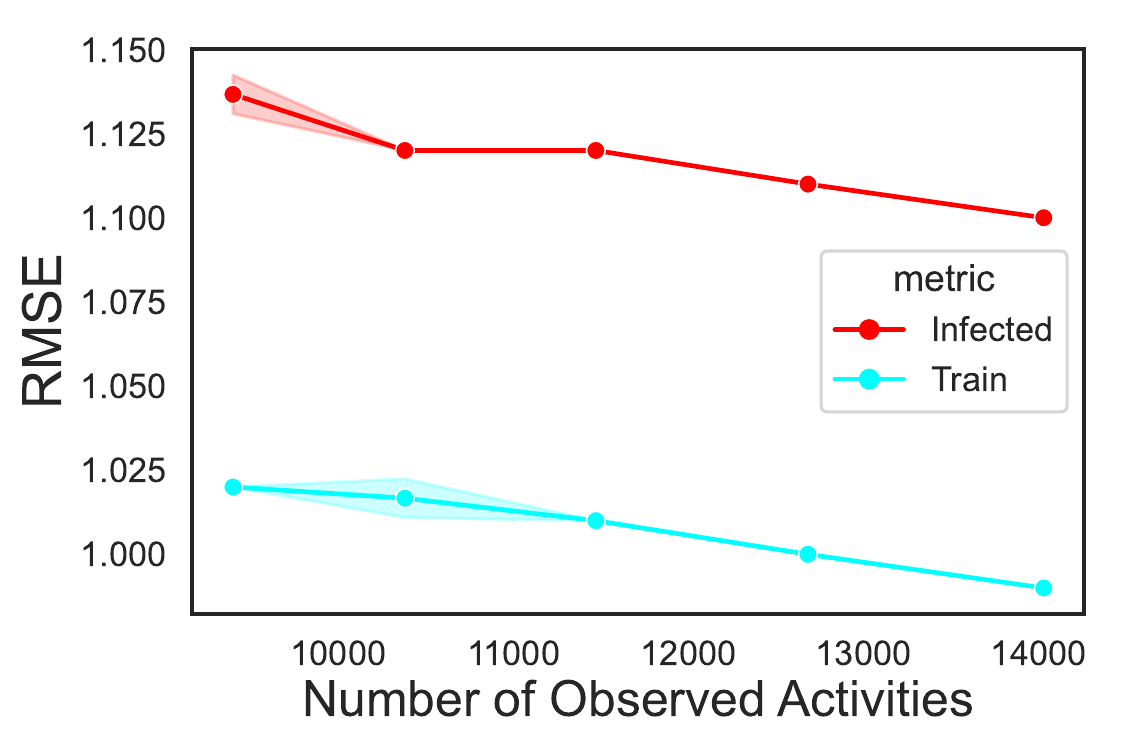}
			\label{fig:casremovestrain}
		\end{subfigure}
		\caption{{  Impact of cascade removal rate on the infection time of diffusion matrix results for LFR100 dataset.
		}}
		\label{fig:removecascade}
	\end{center}
\end{figure}

\noindent(2) How does the density of the ground-truth network structure affects the inference of missing information?\\ We simulated four different LFR datasets with $50$ nodes and $50$ generated cascades with different {mixing parameters} to have various densities.
Then in each dataset, we randomly removed the activities in cascades with 0.3 rates and the remaining $60\%$ of links as observed data. The results are reported based on an average of five training sets and their standard deviations. The sampling set had $2000$ iterations, burn-in of $1500$, and thinning of $5$. Fig. \eqref{fig:densitystru} shows the ground-truth graph's structural metrics against different densities. As illustrated, increasing the amount of information improves the method's performance. As shown in Fig. \eqref{fig:densitycas}, more links lead to better cascade estimation. Moreover, we achieve acceptable results only by observing approximately half of the data in structure and diffusion layers. 
\\(3) How is the performance of DiffStru even if there is no activity observation for a subset of users (e.g., users with private accounts)?\\ In the LFR50 dataset, we used the same removal setting as the previous analysis mentioned in question (2). We increased the amount of data deletion by removing the complete information of five rows of diffusion matrix to simulate the private user profiles. In addition to accessing half of the user's activities, we did not have any data about the activities of five network nodes. Then, two settings of the proposed method (correlated and independent priors) were tested. Fig. \eqref{fig:privateuser} demonstrates the result of DiffStru when the hyper-parameter covariance matrix of the prior distribution is initialized with correlated side information and when the priors are independent by utilizing an identity matrix. We found that the dependent prior can increase the performance of DiffStru for both structure and diffusion information, especially when there are empty rows in the cascade matrix.
\\(4) Can the learned latent factors be used for classification problems?\\ Each node and cascade can be represented with an embedded $D \times 1$ vector.
LFR100 and LFR400 datasets have four and six ground-truth embedded community structures, respectively. We visualized the network nodes from the learned matrix $\mathbf{X}$ and $\mathbf{U}$ with the color of their communities as shown in Fig. \eqref{fig:LFR100-X}, \eqref{fig:LFR100-U}, \eqref{fig:LFR400-X}, and \eqref{fig:LFR400-U}, by using the PCA and t-SNE methods \cite{tsne2008visualizing}. The figure illustrates that nodes' embedded learned features are separated in space according to their community labels. At the same time, we did not consider any assumptions about the community structure in DiffStru. Despite node classification, we do not have any ground truth for comparing the cascade classifications. Still, their embedded vectors based on the learned matrix $\mathbf{Y}$ are shown in Fig. \eqref{fig:LFR100-Y} and \eqref{fig:LFR400-Y}.
\\(5) What is the impact of the removal rate on the performance of DiffStru?\\
We investigated this issue on the LFR100 dataset. First, keeping the complete diffusion information, we examined the effect of link removal on the output of link mining. In this scenario, there was no test for cascades. The infection times estimator was tested in the training set of cascades. We randomly chose $203$ samples of links from the structure network for the test. Then in experiments, we decreased the number of observations to find out the impact of link observation on the method's performance, as shown in Fig. \eqref{fig:removelink}. 
Second, we tested the different missing rates in the cascade data using the full observed graph of nodes. We chose $1655$ activities as a test and did the experiments on various observations. Here, we evaluated cascade information, and there were no missing links for testing. In Fig. \eqref{fig:removecascade}, the performance of DiffStru on the two data splits is represented. RMSE on the training data is almost constant while decreasing by increasing the number of observations in the test set. 

\subsection{Comparison }\label{sec:comparison}
We tested DiffStru against related works on two synthetics and two real datasets. Based on the description in Section \eqref{sec:setting}, different missingness patterns were utilized for generating the test scenarios. The Gibbs sampler of DiffStru was run with $5000$ iterations, burn-in of $4500$, and thinning of $5$. { The comparison over the structural network is reflected in Table \eqref{fig:randomnontable} for non-random and random missing.}. In both cases, DiffStru outperforms others in different metrics, especially simultaneously in AUC and F-measure, which are essential performance metrics for imbalanced datasets. { The input for CN, FPMF, AA, RA, and LCPA is the observed network, while Netrate, DANI, and REFINE use the observed cascades. DiffStru and JWNMF utilize both observed networks and cascades}. 

\begin{figure}[t]
	\begin{center}
		\begin{subfigure}{0.4\textwidth}
			\centering\includegraphics[width=1\textwidth]{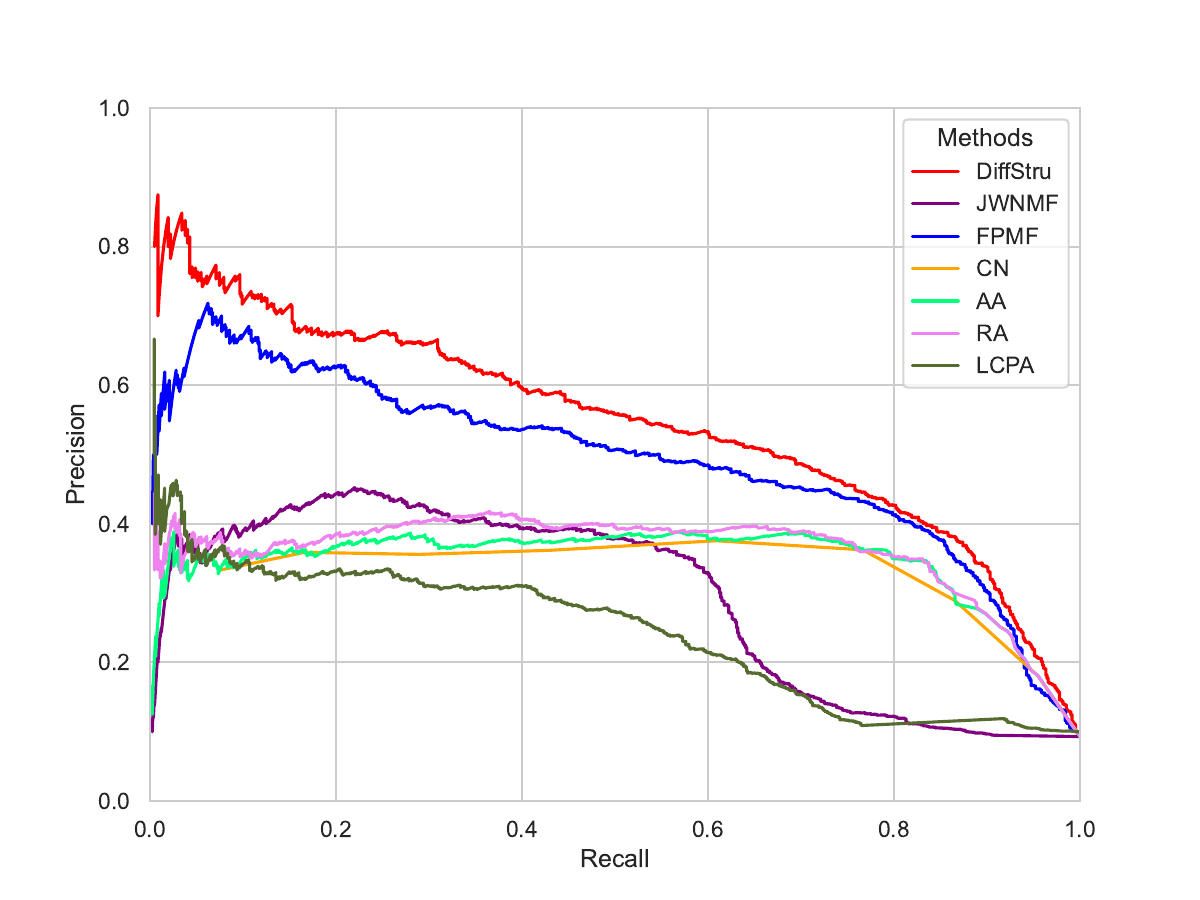}
			\caption{LFR100}
			\label{fig:PRLFR100}
		\end{subfigure}
		\begin{subfigure}{0.4\textwidth}
			\centering\includegraphics[width=1\textwidth]{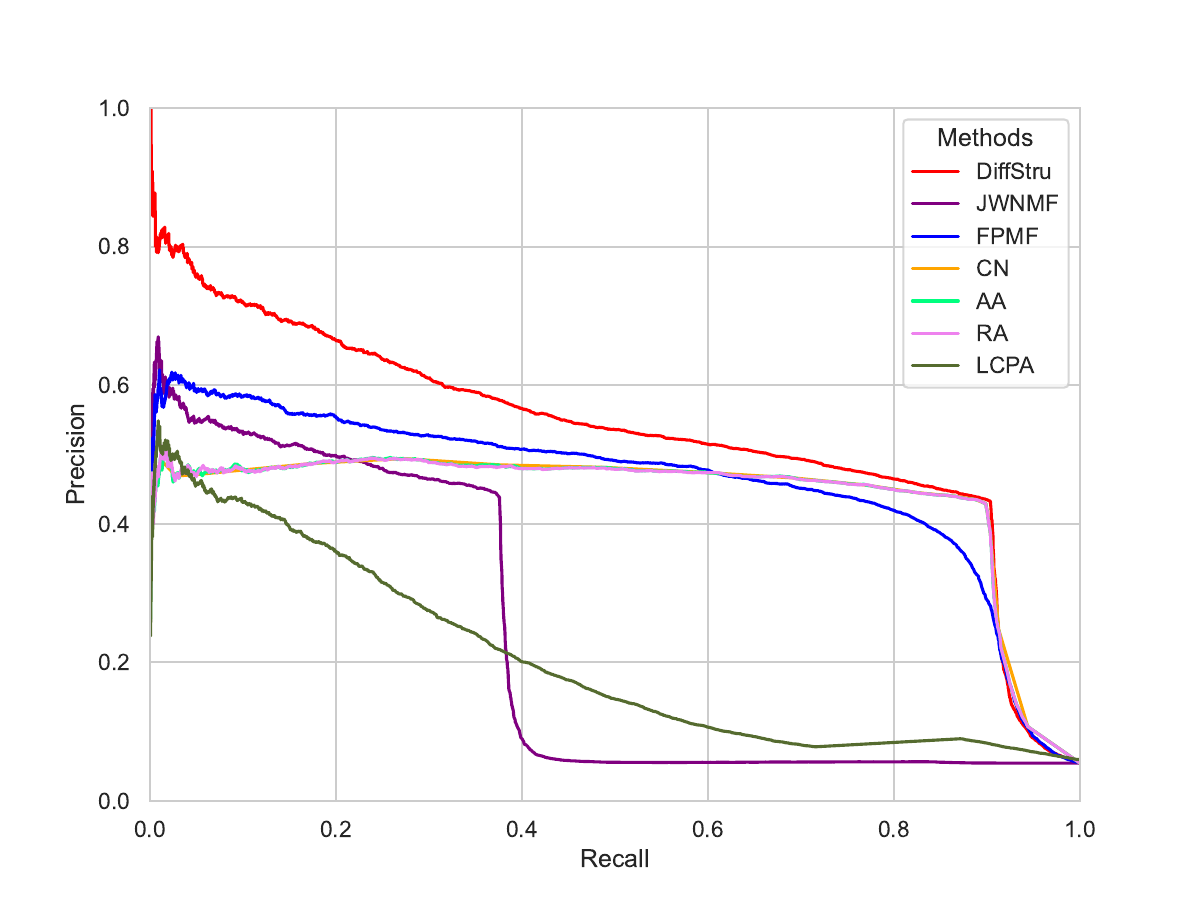}
			\caption{LFR400}
			\label{fig:PRLFR400}
		\end{subfigure}
		\begin{subfigure}{0.4\textwidth}
			\centering\includegraphics[width=1\textwidth]{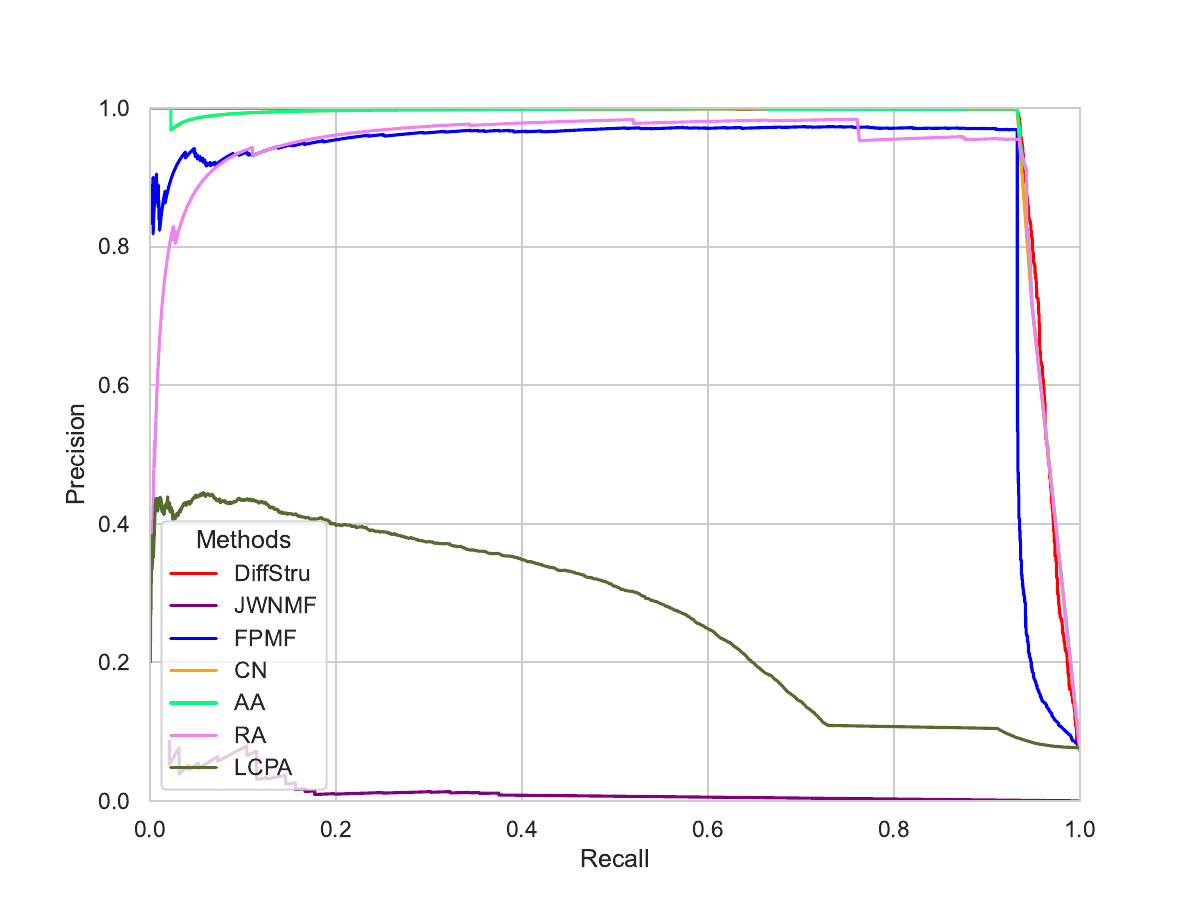}
			\caption{Twitter}
			\label{fig:PRTwitter}
		\end{subfigure}
		\begin{subfigure}{0.4\textwidth}
			\centering\includegraphics[width=1\textwidth]{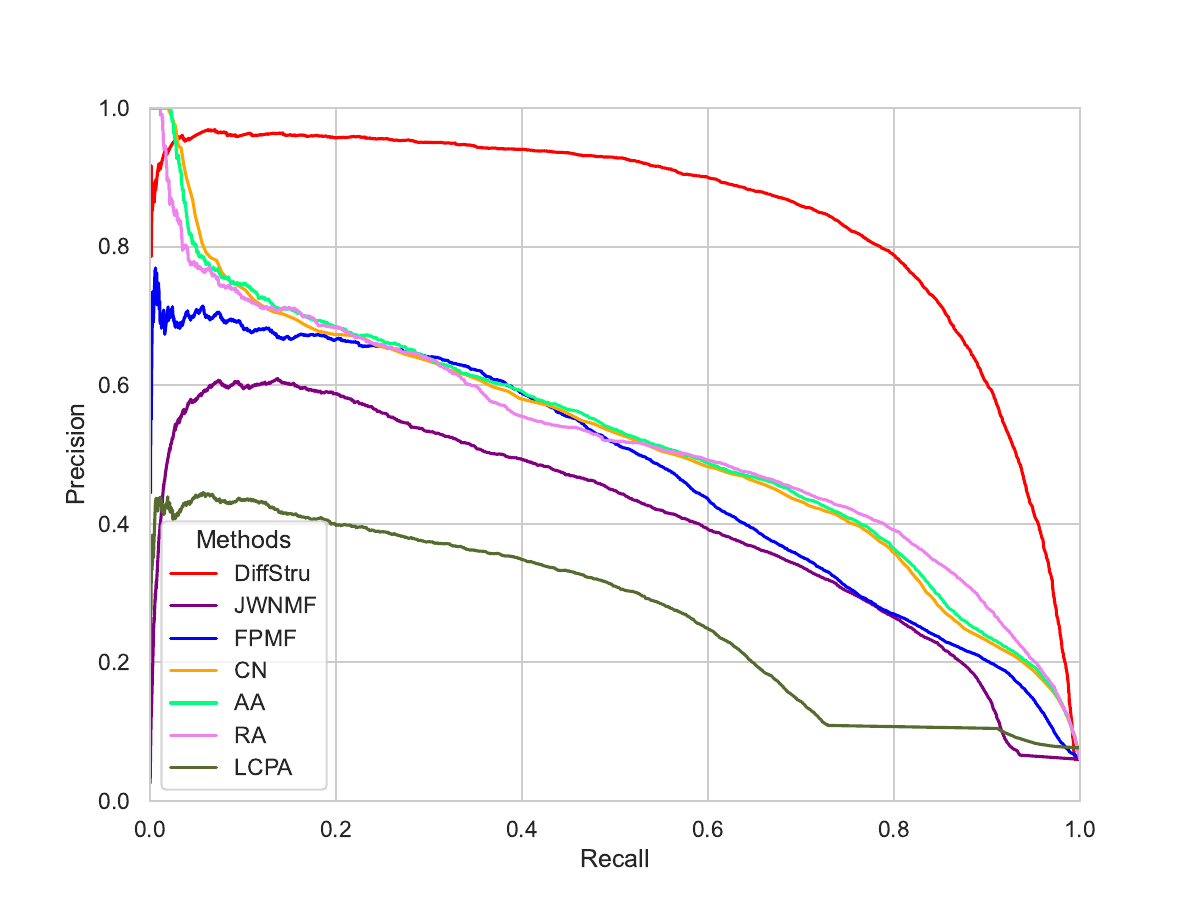}
			\caption{Memetracker}
			\label{fig:PRMemes}
		\end{subfigure}
		\caption{{\color{black} Precision-Recall curve on four different datasets. DiffStru achieves more break-even point value than other methods in completing the structural network.}}
		\label{fig:PRcurve}
	\end{center}
\end{figure}

{\scriptsize
\begin{table}[t]
	\caption
	{{\color{black}
		Comparison results for mining omitted activities of cascades in random and nonrandom deletion.
	}}
	
	\label{fig:cascadetableforall}
	\centering
	\begin{tabular}{| c| c| c| c| c| c| c|}
		\hline
		& Dataset                  & Metric            & DiffStru & DeepDiff & Reg-1 & Reg-2 \\ \hline
		\parbox[t]{4pt}{\multirow{8}{*}{\rotatebox[origin=c]{90}{Random}}}     & \multirow{2}{*}{LFR100}  & RMSE              & 0.97     & 0.22     & 0.47  & 0.47     \\ \cline{3-7}
		&                          & MAP@10,50,100 & - & 7.90, 10.20, 10.60 & $\times$ & $\times$ \\ \cline{2-7}
		& \multirow{2}{*}{LFR400}  & RMSE              & 0.47 & 0.05 & 0.26 & 0.24 \\  \cline{3-7}
		&                          & MAP@10,50,100 & - & 1.11, 1.60, 1.90 & $\times$  & $\times$ \\ \cline{2-7}
		& \multirow{2}{*}{Twitter} & RMSE              & 0.11 & 0.18 & 0.16 & 0.16 \\ \cline{3-7}
		&                          & MAP@10,50,100 & - & 12.70, 15.30, 15.40 & $\times$ &  $\times$ \\ \cline{2-7}
		& \multirow{2}{*}{Mem}     & RMSE              & 1.79 & 1.99 & 1.98 & 2.55 \\ \cline{3-7}
		&                          & MAP@10,50,100 & - & 8.30, 10.04, 10.70 & $\times$ & $\times$ \\ \hline
		\parbox[t]{4pt}{\multirow{8}{*}{\rotatebox[origin=c]{90}{Non-random}}} & \multirow{2}{*}{LFR100}  & RMSE              & 0.98 & 0.13 & 0.45 & 0.39 \\ \cline{3-7}
		&                          & MAP@10,50,100 & - & 11.10, 13.20, 13.50 & $\times$ &  $\times$ \\ \cline{2-7}
		& \multirow{2}{*}{LFR400}  & RMSE              & 0.49 & 0.03 & 0.26 & 0.23 \\ \cline{3-7}
		&                          & MAP@10,50,100 & - & 1.30, 1.80, 2.00 & $\times$ & $\times$ \\ \cline{2-7}
		& \multirow{2}{*}{Twitter} & RMSE              & 0.10 & 0.07 & 0.06 & 0.08 \\ \cline{3-7}
		&                          & MAP@10,50,100 & - & 5.70, 6.80, 7.30 & $\times$ & $\times$ \\ \cline{2-7}
		& \multirow{2}{*}{Mem}     & RMSE             & 3.40 & 2.66 & 2.41 & 2.97 \\ \cline{3-7}
		&                          & MAP@10,50,100 & - & 10.20, 12.00, 12.30  & $\times$ & $\times$ \\ \hline
		
	\end{tabular}
\end{table}
}

\begin{figure}[h]
	\begin{center}
		\begin{subfigure}{0.4\textwidth}
			\centering\includegraphics[width=1\textwidth]{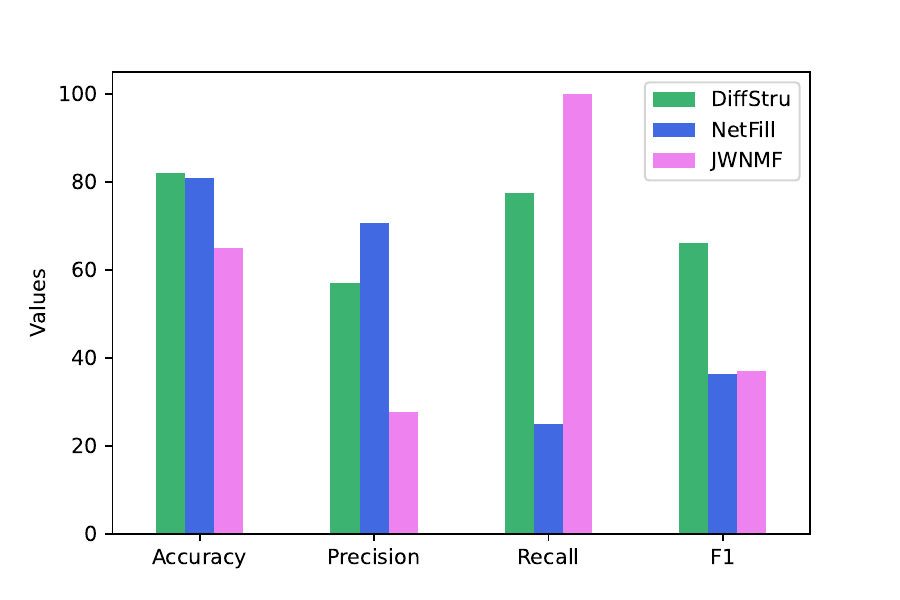}
			\caption{LFR100}
			\label{fig:NetFillCompareLFR}
		\end{subfigure}
		\begin{subfigure}{0.4\textwidth}
			\centering\includegraphics[width=1\textwidth]{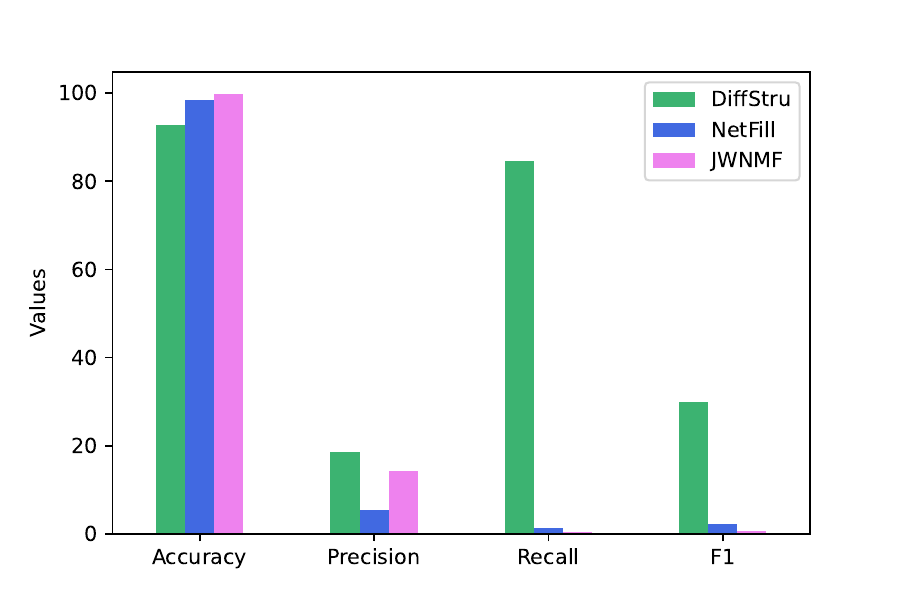}
			\caption{Twitter}
			\label{fig:NetFillCompareTwitter}
		\end{subfigure}
		\caption{\color{black} Detection of missing nodes in cascades without inferring the infection time.}
		\label{fig:NetFillCompare}
	\end{center}
\end{figure}

Analysis from the diffusion perceptive is reported in Table \eqref{fig:cascadetableforall}. 
Notably, the range of RMSE value does not have a limited ceiling. The remarkable point is that DiffStru infers any missing cascade node with its infection time. Still, DeepDiffuse outputs a unique infection time and sorts all network nodes according to the probability suggested for the next infected node in the cascade sequence. While DiffStru exactly infers the missing node and its timestamp, one should evaluate the precision of DeepDiffuse for discovering the node. Besides, the output of DiffStru is a single node with a real-value time, while DeepDiffuse ranks the nodes. We also had to report the MAP@K metric for DeepDiffuse, while this metric is not necessary for DiffStru.\\
For the structural network, the ratio of unlinked pairs against the linked nodes is too high; hence, we face the problem of imbalanced data. AUC and F-measure are the popular metrics for comparing the imbalanced classification. In all cases of random and non-random missing in four different synthetic and real datasets, DiffStru outperforms the other methods in terms of AUC. AUC shows that the overall performance of DiffStru at all different thresholds is almost high. For real-world data, to have better interpretability, we should compare methods in terms of F-measure by choosing the best threshold for each classifier to explicitly assign samples to two different classes. DiffStru is better than all the competing methods in terms of F-measure, except for the Twitter dataset, in which the CN method has a slightly better F-measure. The same pattern occurs in the accuracy metric. However, accuracy is not a good metric for imbalanced applications. Moreover, the higher SRE values indicate that the oracle adjacency matrix is recovered with less error. On average, the performance of DiffStru is high for the SRE criteria. For the MCC metric, the probabilistic methods (first DiffStru, and then FPMF) have the best performance compared to the classical methods. 
Finally, the precision-recall curve for the competing methods is shown in Fig. \eqref{fig:PRcurve}. The break-even points of DiffStru are near $0.54$,$0.53$,$0.94$, and $0.79$ for LFR100, LFR400, Twitter, and Memetracker datasets, respectively, which are all higher than the other methods. In general, DiffStru performs better than the competing methods regarding different metrics of Section \eqref{sec:metrics}.
Based on the results, REFINE from the network inference category performed less well than other works because it is more sensitive to missing data, a factor it does not consider.\\
For the diffusion network comparisons, due to the lack of similar work, we had to make unfair comparisons with DeepDiffuse and Regression models for evaluating the RMSE of missing infection times. DeepDiffuse provides a sorted list of nodes instead of suggesting a specific missed node. The RMSE of infection times for the test data and MAP@K for predicting the name of a node are listed in Table \eqref{fig:cascadetableforall}. Since the regression models cannot detect the node's identity, the corresponding table cells for MAP@K are filled with $\times$, and just RMSE is reported for its two versions, Reg-1 and Reg-2. On the other hand, DiffStru finds the missing node with its infection time, and its RSME is reported in the table. Because of the exact inference of DiffStru, its MAP@K is always $100\%$, and hence we have reported this fact by filling the corresponding cells in the table with $-$. It can be seen that despite the accurate identification of the missing node with DiffStru, its RMSE for infection time is in the appropriate range. In the RMSE comparison, it is important to note that the maximum value of RMSE is not limited to one. {\color{black} Fig. \eqref{fig:NetFillCompare} illustrates the comparison between our proposed method, NetFill, and JWNMF. The JWNMF and NetFill algorithms attempt to determine the list of missing nodes without inferring the time of infection. Due to the incompleteness of the network structure, NetFill and JWNMF perform worse than DiffStru. A total number of $636$ cascades participating nodes are missed in Twitter, while Diffstru detects $539$, NetFill detects $9$, and JWNMF detects $835$. Among the $2954$ cascades participating nodes missing for LFR100, DiffStru, NetFill, and JWNMF inferred $2202$, $705$, and $7934$.}

\begin{table}
	\centering
		\caption{ Link prediction result for random and non-random deletion.}
	\label{fig:randomnontable}
	\includegraphics[width=1\textwidth]{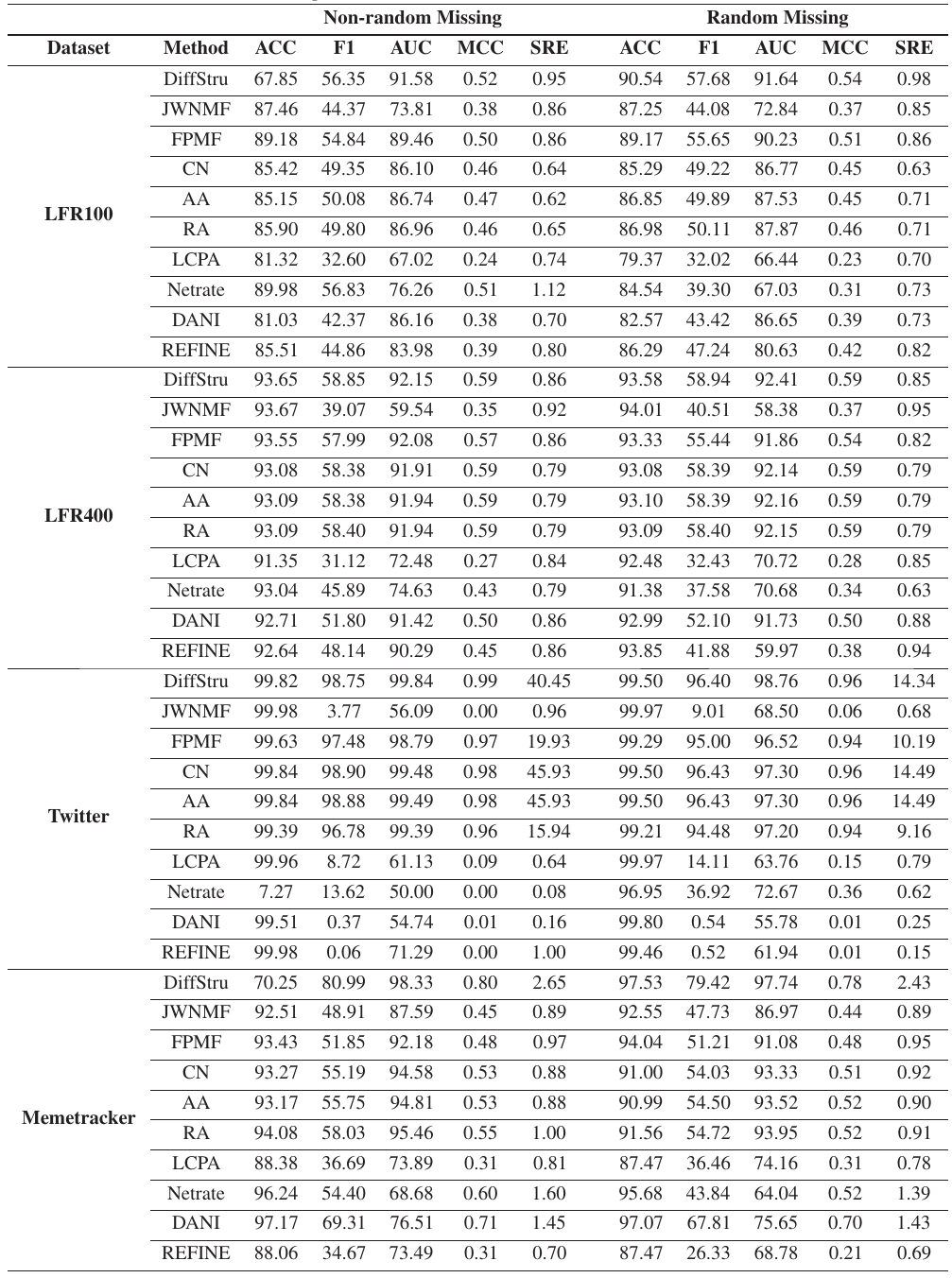}

\end{table}
\newpage
\section{Conclusion and Discussion}\label{sec:conclusion}

Although, in the past, researchers have worked on predicting the missing links of the network using existing links or inferring the structure from the cascades, completing cascades and links with the concurrent incompleteness in both of them has not been considered.
In this paper, we presented a novel generative model called DiffStru by combing a network's partially observed structure and diffusion information to infer the missing data efficiently. We embedded the observations in a low-dimensional latent space by fitting suitable distributions and estimating parameters with Gibbs sampling. DiffStru learns the latent factors during inference and can be utilized to solve other related network classification problems, such as community detection.  We conducted several experiments on synthetic and real datasets to measure the effectiveness of DiffStru. 
{To extend the proposed method, we can use different prior distributions for different link observer variables if we can obtain more network data, such as the profile and features of users.} {\color{black}In future work, instead of the inner product of latent matrices, a deep-learning approach can be utilized for model inference to capture the complex relations of the data. In addition, our generative model can be extended to support more coupling relations.} {Another future direction of this work is extending the model to a dynamic framework by using the diffusion and structure of previous snapshots to forecast the next step of network evolution and predict the cascade sequences. Therefore, the model can infer the addition and removal of links caused by users joining or leaving the network. The creation time of links can also be estimated.}

\newpage

\bibliographystyle{unsrtnat}
\bibliography{references}  






\end{document}